\documentclass[12pt]{article}

\usepackage{graphicx}
\usepackage{float}
\usepackage{cite}
\usepackage{amsfonts}
\usepackage{amssymb}
\usepackage{xcolor}

\def\gtwid{\mathrel{\raise.3ex\hbox{$>$\kern-.75em\lower1ex\hbox{$\sim$}}}}
\def\ltwid{\mathrel{\raise.3ex\hbox{$<$\kern-.75em\lower1ex\hbox{$\sim$}}}}
\def\square{\kern1pt\vbox{\hrule height 1.2pt\hbox{\vrule width 1.2pt\hskip 3pt
   \vbox{\vskip 6pt}\hskip 3pt\vrule width 0.6pt}\hrule height 0.6pt}\kern1pt}

\begin{document}

\begin{titlepage}

\begin{flushright}
UFIFT-QG-21-01 , CCTP-2020-4
\end{flushright}

\vskip 1cm

\begin{center}
{\bf Graviton Self-Energy from Gravitons in Cosmology}
\end{center}

\vskip .5cm

\begin{center}
L. Tan$^{1\star}$, N. C. Tsamis$^{2\dagger}$ and R. P. Woodard$^{1\ddagger}$
\end{center}

\vskip .5cm

\begin{center}
\it{$^{1}$ Department of Physics, University of Florida,\\
Gainesville, FL 32611, UNITED STATES}
\end{center}

\begin{center}
\it{$^{2}$ Institute of Theoretical Physics \& Computational Physics, \\
Department of Physics, University of Crete, \\
GR-710 03 Heraklion, HELLAS}
\end{center}

\vspace{0.5cm}

\begin{center}
ABSTRACT
\end{center}
Although matter contributions to the graviton self-energy 
$-i[\mbox{}^{\mu\nu} \Sigma^{\rho\sigma}](x;x')$ must be separately
conserved on $x^{\mu}$ and ${x'}^{\mu}$, graviton contributions obey
the weaker constraint of the Ward identity, which involves a divergence
on both coordinates. On a general homogeneous and isotropic background
this leads to just four structure functions for matter contributions but
nine structure functions for graviton contributions. We propose a convenient
parameterization for these nine structure functions. We also apply the 
formalism to explicit one loop computations of $-i[\mbox{}^{\mu\nu} 
\Sigma^{\rho\sigma}](x;x')$ on de Sitter background, one of the 
contributions from a massless, minimally coupled scalar and the other for
the contribution from gravitons in the simplest gauge. We also specialize 
the linearized, quantum-corrected Einstein equation to the graviton mode
function and to the gravitational response to a point mass.

\begin{flushleft}
PACS numbers: 04.50.Kd, 95.35.+d, 98.62.-g
\end{flushleft}

\vskip .5cm

\begin{flushleft}
{\it This paper is dedicated to Stanley Deser on the occasion of his 90th 
birthday.}
\end{flushleft}

\vskip .5cm

\begin{flushleft}
$^{\star}$ e-mail: ltan@ufl.edu \\
$^{\dagger}$ e-mail: tsamis@physics.uoc.gr \\
$^{\ddagger}$ e-mail: woodard@phys.ufl.edu
\end{flushleft}

\end{titlepage}

\section{Introduction}

The graviton self-energy $-i[\mbox{}^{\mu\nu} \Sigma^{\rho\sigma}](x;x')$ is the 
1PI (one-particle-irreducible) 2-graviton function. It can be used to quantum-correct 
the linearized Einstein equation,
\begin{equation}
\mathcal{L}^{\mu\nu\rho\sigma} h_{\rho\sigma}(x) - \int \!\! d^4x' \,
\Bigl[\mbox{}^{\mu\nu} \Sigma^{\rho\sigma}\Bigr](x;x') h_{\rho\sigma}(x') = \frac12
\kappa T^{\mu\nu}_{\rm lin}(x) \; , \label{Einsteineqn}
\end{equation}
where $h_{\mu\nu}(x)$ is the graviton field, $\mathcal{L}^{\mu\nu\rho\sigma}$ is the
Lichnerowicz operator in the appropriate background geometry, $T^{\mu\nu}_{\rm lin}$ 
is the linearized stress tensor density and $\kappa^2 \equiv 16 \pi G$ is the loop
counting parameter of quantum gravity. Equation (\ref{Einsteineqn}) can be used to
study how quantum corrections change the propagation of gravitational radiation and 
also how they affect gravitational forces. Note that equation 
(\ref{Einsteineqn}) does not quite represent a semi-classical approach to gravity 
because the graviton self-energy receives contributions from the 0-point fluctuations
of gravity as well as matter.

Quantum corrections on flat space background make no change at all to the kinematics
of gravitons, and corrections to gravitational potentials only become significant at
the Planck length \cite{Radkowski:1970,Capper:1973mv,Capper:1973bk,Capper:1974ed,
Duff:1974ud,Donoghue:1993eb,Dalvit:1994gf,Donoghue:1994dn,Muzinich:1995uj,
Hamber:1995cq,Akhundov:1996jd,Duff:2000mt,Khriplovich:2002bt,BjerrumBohr:2002ks,
BjerrumBohr:2002kt,Khriplovich:2004cx,Satz:2004hf}. However, the situation can be 
very different in cosmology, especially during primordial inflation. Accelerated 
expansion rips light scalars and gravitons from the vacuum, causing secular 
enhancements of the graviton field strength \cite{Mora:2013ypa}, and changes in 
gravitational potentials that grow nonperturbatively strong at large distances and 
at late times \cite{Park:2015kua}.

The purpose of this paper is to develop a technique for representing the graviton
self-energy on a general homogeneous, isotropic and spatially flat background, 
with scale factor $a(\eta)$ and Hubble parameter $H(\eta)$,
\begin{equation}
ds^2 = a^2(\eta) \Bigl[-d\eta^2 + d\vec{x} \!\cdot\! d\vec{x}\Bigr] \qquad 
\Longrightarrow \qquad H(\eta) \equiv \frac{\partial_0 a}{a^2} \; . \label{geometry}
\end{equation}
Our representation consists of a sum of tensor differential operators acting on
four scalar structure functions. Similar representations have been already given 
for matter contributions to the graviton self-energy on de Sitter background 
\cite{Park:2011ww,Leonard:2014zua} but it is cumbersome to infer the structure 
functions from the primitive result, and then use them in the effective field 
equation (\ref{Einsteineqn}). Another problem is that graviton contributions to the 
graviton self-energy require five new structure functions.

To understand the difference between matter contributions and those from gravity
itself, first use general tensor analysis on the background (\ref{geometry}) to
construct 21 basis tensors $[\mbox{}^{\mu\nu} \mathcal{D}^{\rho\sigma}]$ from 
$\delta^{\mu}_{~0}$, the spatial part of the Minkowski metric $\overline{\eta}^{
\mu\nu} \equiv \eta^{\mu\nu} + \delta^{\mu}_{~0} \delta^{\nu}_{~0}$ and the 
spatial derivative operator $\overline{\partial}^{\mu} \equiv \partial^{\mu} +
\delta^{\mu}_{~0} \partial_0$. The graviton self-energy can be expressed as a
sum of these operators acting on scalar functions of $\eta$, $\eta'$ and
$\Vert \vec{x} - \vec{x}'\Vert$,
\begin{equation}
-i\Bigl[\mbox{}^{\mu\nu} \Sigma^{\rho\sigma}\Bigr](x;x') = \sum_{i=1}^{21}
\Bigl[\mbox{}^{\mu\nu} \mathcal{D}_{i}^{\rho\sigma}\Bigr] \!\times\! T^i(x;x') 
\; . \label{initialrep}
\end{equation}
The 21 basis tensors are listed in Table~\ref{Tbasis}.
\begin{table}[H]
\setlength{\tabcolsep}{8pt}
\def\arraystretch{1.5}
\centering
\begin{tabular}{|@{\hskip 1mm }c@{\hskip 1mm }||c||c|c||c|c|}
\hline
$i$ & $[\mbox{}^{\mu\nu} \mathcal{D}^{\rho\sigma}_i]$ & $i$ & $[\mbox{}^{\mu\nu} 
\mathcal{D}^{\rho\sigma}_i]$ & $i$ & $[\mbox{}^{\mu\nu} \mathcal{D}^{\rho\sigma}_i]$ \\
\hline\hline
1 & $\overline{\eta}^{\mu\nu} \overline{\eta}^{\rho\sigma}$ & 8 & $\overline{\partial}^{\mu} 
\overline{\partial}^{\nu} \overline{\eta}^{\rho\sigma}$ & 15 & $\delta^{(\mu}_{~~0} 
\overline{\partial}^{\nu)} \delta^{\rho}_{~0} \delta^{\sigma}_{~0}$ \\
\hline
2 & $\overline{\eta}^{\mu (\rho} \overline{\eta}^{\sigma) \nu}$ & 9 & $\delta^{(\mu}_{~~0} 
\overline{\eta}^{\nu) (\rho} \delta^{\sigma)}_{~~0}$ & 16 & $\delta^{\mu}_{~0} \delta^{\nu}_{~0} 
\overline{\partial}^{\rho} \overline{\partial}^{\sigma}$ \\
\hline
3 & $\overline{\eta}^{\mu\nu} \delta^{\rho}_{~0} \delta^{\sigma}_{~0}$ & 10 & 
$\delta^{(\mu}_{~~0} \overline{\eta}^{\nu) (\rho} \overline{\partial}^{\sigma)}$ & 17 & 
$\overline{\partial}^{\mu} \overline{\partial}^{\nu} \delta^{\rho}_{~0} \delta^{\sigma}_{~0}$ \\
\hline
4 & $\delta^{\mu}_{~0} \delta^{\nu}_{~0} \overline{\eta}^{\rho\sigma}$ & 11 & 
$\overline{\partial}^{(\mu} \overline{\eta}^{\nu) (\rho} \delta^{\sigma)}_{~~0}$ & 18 
& $\delta^{(\mu}_{~~0} \overline{\partial}^{\nu)} \delta^{(\rho}_{~~0} 
\overline{\partial}^{\sigma)}$ \\
\hline
5 & $\overline{\eta}^{\mu\nu} \delta^{(\rho}_{~~0} \overline{\partial}^{\sigma)}$ & 12 & 
$\overline{\partial}^{(\mu} \overline{\eta}^{\nu)(\rho} \overline{\partial}^{\sigma)}$ & 
19 & $\delta^{(\mu}_{~~0} \overline{\partial}^{\nu)} \overline{\partial}^{\rho} 
\overline{\partial}^{\sigma}$ \\
\hline
6 & $\delta^{(\mu}_{~~0} \overline{\partial}^{\nu)} \overline{\eta}^{\rho\sigma}$ & 13 & 
$\delta^{\mu}_{~0} \delta^{\nu}_{~0} \delta^{\rho}_{~0} \delta^{\sigma}_{~0}$ & 20 & 
$\overline{\partial}^{\mu} \overline{\partial}^{\nu} \delta^{(\rho}_{~~0} 
\overline{\partial}^{\sigma)}$ \\
\hline
7 & $\overline{\eta}^{\mu\nu} \overline{\partial}^{\rho} \overline{\partial}^{\sigma}$ & 
14 & $\delta^{\mu}_{~0} \delta^{\nu}_{~0} \delta^{(\rho}_{~~0} \overline{\partial}^{\sigma)}$ 
& 21 & $\overline{\partial}^{\mu} \overline{\partial}^{\nu} \overline{\partial}^{\rho} 
\overline{\partial}^{\sigma}$ \\
\hline
\end{tabular}
\caption{\footnotesize The 21 basis tensors used in expression (\ref{initialrep}). 
Note that the pairs $(3,4)$, $(5,6)$, $(7,8)$, $(10,11)$, $(14,15)$, $(16,17)$ and 
$(19,20)$ are related by reflection.}
\label{Tbasis}
\end{table}
\noindent Now note that 7 of the scalar coefficient functions are related by 
reflection invariance,
\begin{equation}
-i \Bigl[\mbox{}^{\mu\nu} \Sigma^{\rho\sigma}\Bigr](x;x') =
-i \Bigl[\mbox{}^{\rho\sigma} \Sigma^{\mu\nu}\Bigr](x';x) \; . \label{reflection}
\end{equation}
The various relations are listed in Table~\ref{ReflectionT}.
\begin{table}[H]
\setlength{\tabcolsep}{8pt}
\def\arraystretch{1.5}
\centering
\begin{tabular}{|@{\hskip 1mm }c@{\hskip 1mm }||c||c|c|}
\hline
$i$ & Relation & $i$ & Relation \\
\hline\hline
$3,4$ & $T^4(x;x') = +T^3(x';x)$ & $14,15$ & $T^{15}(x;x') = -T^{14}(x';x)$ \\
\hline
$5,6$ & $T^6(x;x') = -T^5(x';x)$ & $16,17$ & $T^{17}(x;x') = +T^{16}(x';x)$ \\
\hline
$7,8$ & $T^8(x;x') = +T^7(x';x)$ & $19,20$ & $T^{20}(x;x') = -T^{19}(x';x)$ \\
\hline
$10,11$ & 
$T^{11}(x;x') = -T^{10}(x';x)$ & $$ & $$ \\
\hline
\end{tabular}
\caption{\footnotesize Scalar coefficient functions in expression (\ref{initialrep})
which are related by reflection.}
\label{ReflectionT}
\end{table}

The 14 algebraically independent scalar coefficient functions $T^i(x;x')$ are
related by differential equations whose number depends upon whether the
contributions to $-i[\mbox{}^{\mu\nu} \Sigma^{\rho\sigma}](x;x')$ come from
matter or from gravity itself. To understand these relations it is useful to 
define the {\it Ward Operator}, 
\begin{equation}
\mathcal{W}^{\mu}_{~\alpha\beta}(x) \equiv \delta^{\mu}_{~(\alpha} \partial_{\beta)}
+ H a \delta^{\mu}_{~0} \eta_{\alpha\beta} \; . \label{Wardop}
\end{equation}
Because matter interacts with gravity through its conserved stress tensor,
matter contributions to the graviton self-energy must be annihilated by the
Ward operator acting on either point,
\begin{equation}
0 = \mathcal{W}^{\mu}_{~\alpha\beta}(x) \times -i \Bigl[\mbox{}^{\alpha\beta} 
\Sigma^{\rho\sigma} \Bigr](x;x') = 0 = \sum_{i=1}^{10} \Bigl[\mbox{}^{\mu}
\mathcal{D}^{\rho\sigma}\Bigr] \times S^i(x;x') \; . \label{selfmatter}
\end{equation}
The 10 independent tensor factors $[\mbox{}^{\mu} \mathcal{D}^{\rho\sigma}]$
are listed in Table~\ref{Sbasis}.
\begin{table}[H]
\setlength{\tabcolsep}{8pt}
\def\arraystretch{1.5}
\centering
\begin{tabular}{|@{\hskip 1mm }c@{\hskip 1mm }||c||c|c|}
\hline
$i$ & $[\mbox{}^{\mu} \mathcal{D}_i^{\rho\sigma}]$ & $i$ & 
$[\mbox{}^{\mu} \mathcal{D}_i^{\rho\sigma}]$ \\
\hline\hline
$1$ & $\delta^{\mu}_{~0} \delta^{\rho}_{~0} \delta^{\sigma}_{~0}$ 
& $6$ & $2 \overline{\eta}^{\mu (\rho} \delta^{\sigma)}_{~~0}$ \\
\hline
$2$ & $2 \delta^{\mu}_{~0} \delta^{(\rho}_{~~0} \overline{\partial}^{\sigma)}$ 
& $7$ & $2 \overline{\partial}^{\mu} \overline{\partial}^{(\rho} \delta^{\sigma)}_{~~0}$ \\
\hline
$3$ & $\delta^{\mu}_{~0} \overline{\eta}^{\rho\sigma}$ 
& $8$ & $2 \overline{\eta}^{\mu (\rho} \overline{\partial}^{\sigma)}$ \\
\hline
$4$ & $\delta^{\mu}_{~0} \overline{\partial}^{\rho} \overline{\partial}^{\sigma}$ 
& $9$ & $\overline{\partial}^{\mu} \overline{\eta}^{\rho\sigma}$ \\
\hline
$5$ & $\overline{\partial}^{\mu} \delta^{\rho}_{~0} \delta^{\sigma}_{~0}$ 
& $10$ & $\overline{\partial}^{\mu} \overline{\partial}^{\rho} 
\overline{\partial}^{\sigma}$ \\
\hline
\end{tabular}
\caption{\footnotesize Scalar coefficient functions in expression (\ref{initialrep})
which are related by reflection.}
\label{Sbasis}
\end{table}
\noindent From (\ref{selfmatter}) we see that matter contributions to the graviton
self-energy are characterized by $14 - 10 = 4$ independent structure functions.
Gravity does not interact with itself through a conserved vertex. Hence graviton 
contributions to the graviton self-energy obey the weaker condition that they are 
annihilated by acting the Ward operator on {\it both} points,
\begin{eqnarray}
\lefteqn{0 = \mathcal{W}^{\mu}_{~\alpha\beta}(x) \!\times\! 
\mathcal{W}^{\rho}_{~\gamma\delta}(x') \!\times\! -i \Bigl[\mbox{}^{\alpha\beta} 
\Sigma^{\gamma\delta}\Bigr](x;x') = \delta^{\mu}_{~0} \delta^{\rho}_{~0} 
\!\times\! R^1(x;x') + \overline{\eta}^{\mu\rho} } \nonumber \\
& & \hspace{0cm} \times R^2(x;x') 
+ \delta^{\mu}_{~0} \overline{\partial}^{\rho} \!\times\! R^3(x;x')
+ \overline{\partial}^{\mu} \delta^{\rho}_{~0} \!\times\! R^4(x;x')
+ \overline{\partial}^{\mu} \overline{\partial}^{\rho} \!\times\! R^5(x;x') \; .
\qquad \label{selfgravity}
\end{eqnarray}
Because expression (\ref{selfgravity}) involves 5 independent tensors we see that
graviton contributions to $-i[\mbox{}^{\mu\nu} \Sigma^{\rho\sigma}](x;x')$ require
$14 - 5 = 9$ structure functions. Our purpose is to propose a convenient 
representation for these structure functions and to elucidate their role in the 
effective field equation (\ref{Einsteineqn}).

Section 2 derives insights from the vacuum polarization, on flat space and in
cosmology, and from the graviton self-energy on flat space. Our representation is
given in section 3. We also work out the equations for quantum corrections to the
graviton mode function, and for the two potentials that describe the response to
a point mass. Section 4 derives explicit results on de Sitter background for a 
dimensionally regulated computation of the contribution from a massless, minimally 
coupled (MMC) scalar \cite{Park:2011ww}, and for a $D=4$ computation of the 
contribution from gravitons away from coincidence ($x^{\mu} \neq {x'}^{\mu}$)
\cite{Tsamis:1996qk}. Section 5 discusses how to extend the $D=4$ computation to a
fully renormalized result. Our conclusions comprise section 6.

\section{Other Bi-Tensor 1PI 2-Point Functions}

The purpose of this section is to motivate our representation for the graviton
self-energy in cosmology by reviewing simpler bi-tensor 1PI 2-point functions
and simpler backgrounds. The section begins with the vacuum polarization on flat 
space background. We then turn to the graviton self-energy on flat space 
background. The section concludes with the vacuum polarization on a general 
cosmological background (\ref{geometry}).

\subsection{Vacuum Polarization on Flat Space}

The 1PI 2-photon function $i[\mbox{}^{\mu} \Pi^{\rho}](x;x')$ has the evocative name,
``vacuum polarization''. A cumbersome and foolish way of expressing it would be to
give all $4^2 = 16$ of its tensor components as functions of the two points $x^{\mu}$ 
and ${x'}^{\mu}$. A much better way is to consolidate the number of functions by 
using general tensor analysis and reflection invariance. On flat space background 
this results in the form,
\begin{equation}
i\Bigl[\mbox{}^{\mu} \Pi^{\rho}_{\rm flat}\Bigr](x;x') = \eta^{\mu\rho} \!\times\!
A(\Delta x^2) + \partial^{\mu} \partial^{\rho} \!\times\! B(\Delta x^2) \; , 
\label{initialflatPi}
\end{equation}
where the invariant interval in a Feynman propagator is,
\begin{equation}
\Delta x^2 \equiv \Bigl\Vert \vec{x} \!-\! \vec{x}'\Bigr\Vert^2 - \Bigl(\vert 
\eta \!-\! \eta'\vert \!-\! i \epsilon\Bigr)^2 \; . \label{Deltax}
\end{equation}

Because photons couple to a conserved current the vacuum polarization is transverse
on each index,
\begin{equation}
0 = \partial_{\mu} \!\times\! i \Bigl[\mbox{}^{\mu} \Pi^{\rho}_{\rm flat} \Bigr](x;x')
= \partial^{\rho} \Bigl[ A(\Delta x^2) + \partial^2 B(\Delta x^2)\Bigr] \; .
\label{flatPiconservation}
\end{equation}
Conservation implies $A = -\partial^2 B$, which allows us to express the flat space
vacuum polarization in terms of a single structure function,
\begin{equation}
i\Bigl[\mbox{}^{\mu} \Pi^{\rho}_{\rm flat}\Bigr](x;x') = \Bigl[\partial^{\mu} 
\partial^{\rho} - \eta^{\mu\rho} \partial^2 \Bigr] B(\Delta x^2) \equiv
\Pi^{\mu\rho} B(\Delta x^2) \; . \label{finalflatPi}
\end{equation}
It is more usual in the literature of quantum field theory to refer to the structure 
function $B(\Delta x^2)$ by the symbol $i \Pi(\Delta x^2)$.
   
\subsection{Graviton Self-Energy on Flat Space}\label{flatgrav}

The advantages of using structure functions are even greater for the graviton
self-energy. It would be fatuous to express this by giving all $4^4 = 256$ 
components. Just as with the vacuum polarization, it is more efficient to 
exploit symmetries of the background, reflection invariance to express $-i 
[\mbox{}^{\mu\nu} \Sigma^{\rho\sigma}](x;x')$ in terms of five basis tensors,
\begin{eqnarray}
\lefteqn{ -i \Bigl[\mbox{}^{\mu\nu} \Sigma^{\rho\sigma}_{\rm flat}\Bigr](x;x')
= \eta^{\mu\nu} \eta^{\rho\sigma} \!\times\! A(\Delta x^2) + \eta^{\mu (\rho}
\eta^{\sigma) \nu} \!\times\! B(\Delta x^2) + \Bigl[ \eta^{\mu\nu} \partial^{\rho} 
\partial^{\sigma} } \nonumber \\
& & \hspace{0cm}  + \partial^{\mu} \partial^{\nu} \eta^{\rho\sigma}\Bigr] 
\!\times\! C(\Delta x^2) + \partial^{(\mu} \eta^{\nu) (\rho} \partial^{\sigma)} 
\!\times\! D(\Delta x^2) + \partial^{\mu} \partial^{\nu} \partial^{\rho} 
\partial^{\sigma} \!\times\! E(\Delta x^2) \; . \qquad \label{initialflatSigma}
\end{eqnarray}
In this expression and henceforth parenthesized indices are symmetrized, for 
example, $\eta^{\mu (\rho} \eta^{\sigma) \nu} \equiv \frac12 [\eta^{\mu \rho} 
\eta^{\nu\sigma} + \eta^{\mu\sigma} \eta^{\nu\rho}]$. 

\subsubsection{From Matter}

Because matter couples to gravitons through its conserved stress-energy tensor, 
matter contributions to the graviton self-energy must be transverse on each index,
\begin{eqnarray}
\lefteqn{0 = \partial_{\nu} \!\times\! -i \Bigl[\mbox{}^{\mu\nu} 
\Sigma^{\rho\sigma}_{\rm flat}\Bigr](x;x') = \partial^{\mu} \eta^{\rho\sigma}
\Bigl[A \!+\! \partial^2 C\Bigr] } \nonumber \\
& & \hspace{3.5cm} + \eta^{\mu (\rho} \partial^{\sigma)} \Bigl[B \!+\! \frac12 
\partial^2 D \Bigr] + \partial^{\mu} \partial^{\rho} \partial^{\sigma} \Bigl[
C \!+\! \frac12 D \!+\! \partial^2 E\Bigr] \; . \qquad \label{flatSigmamattercons}
\end{eqnarray}
Conservation (\ref{flatSigmamattercons}) allows us to express $A$, $C$ and $E$ in
terms of $C$ and $D$,
\begin{equation}
\Bigl( {\rm Eqn.\ \ref{flatSigmamattercons}}\Bigr) \Longrightarrow A = -\partial^2 
C \quad, \quad B = -\frac12 \partial^2 D \quad , \quad E = -\frac1{\partial^2} 
\Bigl(C \!+\! \frac12 D\Bigr) \; . \label{flatSigmamattercons2}
\end{equation}
Substituting (\ref{flatSigmamattercons2}) in (\ref{initialflatSigma}) results in
another familiar form,
\begin{equation}
\Bigl( {\rm Eqn.\ \ref{flatSigmamattercons}}\Bigr) \Longrightarrow 
-i \Bigl[\mbox{}^{\mu\nu} \Sigma^{\rho\sigma}_{\rm flat}\Bigr](x;x') = \Pi^{\mu\nu}
\Pi^{\rho\sigma} \Bigl(- \frac1{\partial^2} C\Bigr) + \Pi^{\mu (\rho} \Pi^{\sigma) \nu}
\Bigl( -\frac1{2 \partial^2} D\Bigr) \; , \label{finalflatSigmamatter}
\end{equation}
where $\Pi^{\alpha\beta} \equiv \partial^{\alpha} \partial^{\beta} -
\eta^{\alpha\beta} \partial^2$ was introduced in (\ref{finalflatPi}).

\subsubsection{From Gravitons}

Because the couplings of gravitons to themselves are not conserved, the divergence 
on a single index group does not vanish. Of course one can still use general tensor
analysis to parameterize it in terms of three scalar functions,
\begin{equation}
\partial_{\nu} \times -i \Bigl[\mbox{}^{\mu\nu} \Sigma^{\rho\sigma}_{\rm flat}\Bigr](x;x')
= \partial^{\mu} \eta^{\rho\sigma} F(\Delta x^2) + \eta^{\mu (\rho} \partial^{\sigma)} 
G(\Delta x^2) + \partial^{\mu} \partial^{\rho} \partial^{\sigma} \mathcal{H}(\Delta x^2) \; .
\label{flatSigmadivergence}
\end{equation}
The Ward identity requires gravitational contributions to the graviton self energy
to vanish when a divergence is taken on both index groups, 
\begin{equation}
0 = \partial_{\nu} \partial_{\sigma} \!\times\! -i \Bigl[\mbox{}^{\mu\nu} 
\Sigma^{\rho\sigma}_{\rm flat}\Bigr](x;x') = \eta^{\mu\rho} \Bigl[\frac12 \partial^2
G\Bigr] + \partial^{\mu} \partial^{\rho} \Bigl[ F \!+\! \frac12 G \!+\! \partial^2 
\mathcal{H} \Bigr] \; . \label{flatSigmagravitycons1}
\end{equation}
Expression (\ref{flatSigmagravitycons1}) implies,
\begin{equation}
\Bigl( {\rm Eqn.\ \ref{flatSigmagravitycons1}}\Bigr) \qquad \Longrightarrow \qquad 
F = -\partial^2 \mathcal{H} \qquad , \qquad G = 0 \; . 
\label{flatSigmagravitycons2}
\end{equation}

Of course the initial parameterization (\ref{initialflatSigma}) of the graviton 
self-energy pertains to both gravitational and matter contributions. Hence expression
(\ref{flatSigmamattercons}) is still valid for the result of a single divergence.
Comparing (\ref{flatSigmamattercons}) with (\ref{flatSigmadivergence}) allows us
to express the scalar coefficient functions $A$, $B$ and $E$ in terms of $C$, $D$ and 
$\mathcal{H}$,
\begin{equation}
\Bigl( {\rm Eqn.\ \ref{flatSigmagravitycons1}}\Bigr) \Longrightarrow A = -\partial^2 
\Bigl(C \!+\! \mathcal{H} \Bigr) \; , \; B = -
\frac{\partial^2 D}{2} \; , \; E = \frac1{\partial^2} \Bigl(-C \!-\! \frac{D}{2} 
\!+\! \mathcal{H} \Bigr) \; . \label{flatSigmagravitycons3}
\end{equation}
Substituting (\ref{flatSigmagravitycons3}) into (\ref{initialflatSigma}) gives,
\begin{eqnarray}
\lefteqn{ \Bigl( {\rm Eqn.\ \ref{flatSigmagravitycons1}}\Bigr) \qquad \Longrightarrow 
\qquad -i \Bigl[\mbox{}^{\mu\nu} \Sigma^{\rho\sigma}_{\rm flat}\Bigr](x;x') = 
\Pi^{\mu\nu} \Pi^{\rho\sigma} \Bigl( -\frac1{\partial^2} C\Bigr) } \nonumber \\
& & \hspace{2cm} + \Pi^{\mu (\rho} \Pi^{\sigma) \nu} \Bigl( -\frac1{2 \partial^2} 
D\Bigr) + \Bigl[\partial^{\mu} \partial^{\nu} \partial^{\rho} \partial^{\sigma}
\!-\! \eta^{\mu\nu} \eta^{\rho\sigma} \partial^4 \Bigr] \Bigl( \frac1{\partial^2}
\mathcal{H} \Bigr) \; . \qquad \label{finalflatSigmagravity}
\end{eqnarray}

\subsubsection{An Explicit Example}

In section 4 we will reconstruct the structure functions functions from an
explicit computation of $-i[\mbox{}^{\mu\nu} \Sigma^{\rho\sigma}](x;x')$
on de Sitter background \cite{Tsamis:1996qk}. That result was derived using a 
de Sitter breaking gauge in which the graviton propagator consists of three
constant tensor factors, constructed from $\eta^{\mu\nu}$ and $\delta^{\mu}_0$,
which multiply scalar propagators whose expansions in $D = 4$ spacetime 
dimensions have at most two terms \cite{Tsamis:1992xa,Woodard:2004ut}. In 1979
Capper used the flat space limit of this same gauge, with dimensional 
regularization, to compute the one loop contribution to the graviton self-energy 
\cite{Capper:1979ej},
\begin{eqnarray}
\lefteqn{ -i \Bigl[\mbox{}^{\mu\nu} \Sigma_{\rm flat}^{\rho\sigma}\Bigr](x;x') = 
\frac{-\kappa^2}{4 (D^2 \!-\! 1)} \Biggl\{ T_1 \partial^{\mu} \partial^{\nu}
\partial^{\rho} \partial^{\sigma} + T_2 \eta^{\mu\nu} \eta^{\rho\sigma} 
\partial^4 + 2 T_3 \eta^{\mu (\rho} \eta^{\sigma) \nu} \partial^4 }
\nonumber \\
& & \hspace{1cm} T_4 \Bigl[ \eta^{\mu\nu} \partial^2 \partial^{\rho}
\partial^{\sigma} + \partial^{\mu} \partial^{\nu} \eta^{\rho\sigma} \partial^2
\Bigr] + 4 T_5 \partial^{(\mu} \eta^{\nu) (\rho} \partial^{\sigma)} \partial^2
\Biggr\} \Bigl[ i\Delta(x;x')\Bigr]^2 . \quad \label{flatsigma1}
\end{eqnarray}
Here $i\Delta(x;x')$ is the massless propagator in flat space,
\begin{equation}
i\Delta(x;x') = \frac{\Gamma(\frac{D}{2} \!-\! 1)}{4 \pi^{\frac{D}2}} 
\frac1{{\Delta x}^{D-2}} \; . \label{flatprop}
\end{equation}
Capper's results for the coefficients $T_i(D)$ are \cite{Capper:1979ej},
\begin{eqnarray}
T_1(D) & = & \frac{9}{16} D^4 - \frac{21}{16} D^3 - \frac{9}{8} D^2 \; , 
\label{T1} \\
T_2(D) & = & \frac{ \frac{9}{16} D^5 \!-\! \frac{39}{16} D^4 \!-\! \frac{25}{8}
D^3 \!+\! \frac{123}{8} D^2 \!+\! \frac{33}{4} D \!-\! 8}{D \!-\! 2} \; , 
\label{T2} \qquad \\
T_3(D) & = & \frac{ \frac{1}{4} D^4 \!+\! \frac{17}{16} D^3 \!-\! \frac{97}{16} 
D^2 \!-\! \frac{17}{8} D \!+\! 4}{D \!-\! 2} = -T_5(D) \; , \label{T3} \qquad \\
T_4(D) & = & \frac{ -\frac{9}{16} D^5 \!+\! \frac{43}{16} D^4 \!+\! \frac{15}{8}
D^3 \!-\! \frac{119}{8} D^2 \!-\! \frac{25}{4} D \!+\! 8}{D \!-\! 2} \; , 
\label{T4} \qquad
\end{eqnarray}

Comparing expressions (\ref{initialflatSigma}) and (\ref{flatsigma1}) allows
us to identify two of the structure functions,
\begin{equation}
-\frac1{\partial^2} C = \frac{\kappa^2 T_4}{4 (D^2 \!-\! 1)} 
\Bigl[i\Delta^2(x;x')\Bigr]^2 \quad , \quad -\frac1{2 \partial^2} D = 
\frac{\kappa^2 4 T_5}{4 (D^2 \!-\! 1)} \Bigl[i\Delta^2(x;x')\Bigr]^2 \; .
\label{exampleCD}
\end{equation}
The final structure function derives from a comparison of (\ref{flatSigmadivergence})
with the divergence of (\ref{flatsigma1}),
\begin{equation}
\frac1{\partial^2} H = -\frac{\kappa^2 (T_1 \!+\! T_4 \!+\! 2 T_5)}{4 (D^2 \!-\! 1)}
\Bigl[i\Delta^2(x;x')\Bigr]^2 \; . \label{exampleH}
\end{equation}
Substituting (\ref{exampleCD}) and (\ref{exampleH}) in (\ref{finalflatSigmagravity})
gives,
\begin{eqnarray}
\lefteqn{ -i \Bigl[\mbox{}^{\mu\nu} \Sigma_{\rm flat}^{\rho\sigma}\Bigr](x;x') = 
\frac{-\kappa^2}{4 (D^2 \!-\! 1)} \Biggl\{ -T_4 \Pi^{\mu\nu} \Pi^{\rho\sigma}
-2 T_5 \Pi^{\mu (\rho} \Pi^{\sigma) \nu} } \nonumber \\
& & \hspace{2.5cm} + (T_1 \!+\! T_4 \!+\! 2 T_5) \Bigl[ \partial^{\mu} \partial^{\nu} 
\partial^{\rho} \partial^{\sigma} \!-\! \eta^{\mu\nu} \eta^{\rho\sigma} \partial^4
\Bigr] \Biggr\} \Bigl[ i\Delta(x;x')\Bigr]^2 . \qquad \label{flatsigma2}
\end{eqnarray}

Expression (\ref{flatsigma2}) is the dimensionally regulated, primitive contribution.
To renormalize we first isolate ultraviolet divergences using the expansion
\cite{Onemli:2002hr,Onemli:2004mb},
\begin{equation}
\Bigl[ i\Delta(x;x')\Bigr]^2 = \frac{\Gamma(\frac{D}{2} \!-\! 1)}{4 \pi^{\frac{D}2}}
\frac{\mu^{D-4} i\delta^D(x \!-\! x')}{2 (D\!-\!3) (D\!-\!4)} - 
\frac{\partial^2}{32 \pi^4} \Bigl[ \frac{\ln(\mu^2 \Delta x^2)}{\Delta x^2} \Bigr] 
+ O(D \!-\! 4) \; . \label{UVexp}
\end{equation}
At $D=4$ the coefficients $T_4(4) = -\frac{23}{2}$ and $T_5(4) = -\frac{61}{4}$ are
nonzero, but the final coefficient vanishes,
\begin{equation}
T_1(D) \!+\! T_4(D) \!+\! 2 T_5(D) = -\frac14 (D \!-\! 4) D (D \!+\! 1) \; .
\end{equation}
This means that renormalization requires only the Ricci-squared and Weyl-squared 
counterterms. Note also that (\ref{flatsigma2}) must be recovered in the flat space 
limit of the graviton self-energy on de Sitter background \cite{Tsamis:1996qk}. The
final, unregulated result is,
\begin{eqnarray}
\lefteqn{ -i \Bigl[\mbox{}^{\mu\nu} \Sigma_{\rm ren}^{\rho\sigma}\Bigr](x;x') = 
\frac{\kappa^2}{1920 \pi^4} \Bigl[ \frac{23}{2} \Pi^{\mu\nu} \Pi^{\rho\sigma}
+ \frac{61}{2} \Pi^{\mu (\rho} \Pi^{\sigma) \nu} \Bigr] \partial^2 \Bigl[
\frac{\ln(\mu^2 \Delta x^2)}{\Delta x^2} \Bigr]} \nonumber \\
& & \hspace{4.5cm} + \frac{\kappa^2}{96 \pi^2} \Bigl[ \partial^{\mu} \partial^{\nu} 
\partial^{\rho} \partial^{\sigma} \!-\! \eta^{\mu\nu} \eta^{\rho\sigma} \partial^4
\Bigr] i\delta^4(x \!-\! x') \; . \qquad \label{flatsigma3}
\end{eqnarray}
Except for their numerical coefficients, the two nonlocal terms on the first line of 
(\ref{flatsigma3}) could have come from matter contributions; the local term on the 
last line is only possible from gravitational contributions to the graviton
self-energy because its divergence on a single index group is nonzero,
$\partial_{\nu} [\partial^{\mu} \partial^{\nu} \partial^{\rho} \partial^{\sigma} -
\eta^{\mu\nu} \eta^{\rho\sigma} \partial^4] = \partial^{\mu} \partial^2 
\Pi^{\rho\sigma}$. In cosmological backgrounds (\ref{geometry}) we will 
see that distinctly gravitational contributions are much more varied, that they can 
harbor divergences, and that they can be nonlocal.

\subsection{Vacuum Polarization in Cosmology}

Now consider the photon self-energy (aka, the ``vacuum polarization'')
on a general cosmological background (\ref{geometry}).
The symmetries of cosmology are homogeneity and isotropy. This means that the 
initial reduction involves two additional tensors from (\ref{initialflatPi}) and 
that the coefficient functions depend on $\eta$, $\eta'$ and $\Vert \vec{x} - 
\vec{x}'\Vert$. When account is also taken of reflection invariance we can 
write,\footnote{Rather than factors of the spatial gradient 
$\partial^i$, the primitive expression contains one or two factors of the spatial 
coordinate interval $\Delta x^i \equiv x^i - {x'}^i$ multiplied by functions of 
$\Vert \Delta \vec{x}\Vert^2$. These are then written in terms of spatial gradients 
using the identities,
\begin{eqnarray}
\Delta x^i f(\Vert \vec{x}\Vert^2) & = & \frac12 \partial^i I[f] \; , \nonumber \\
\Delta x^i \Delta x^j f(\Vert \vec{x}\Vert^2) & = & \frac14 \partial^i \partial^j
I^2[f] - \frac12 \delta^{ij} I[f] \; , \nonumber
\end{eqnarray}
where $I[f]$ represents the indefinite integral of $f(\Vert \vec{x}\Vert^2)$ with
respect to $\Vert \vec{x}\Vert^2$.}
\begin{eqnarray}
\lefteqn{i \Bigl[\mbox{}^{\mu} \Pi^{\rho}_{\rm cos}\Bigr](x;x') = 
\overline{\eta}^{\mu\rho} A(x;x') + \delta^{\mu}_{~0} \delta^{\rho}_{~0} B(x;x') }
\nonumber \\
& & \hspace{3.5cm} + \delta^{\mu}_{~0} \overline{\partial}^{\rho} C(x;x') -
\overline{\partial}^{\mu} \delta^{\rho}_{~0} C(x';x) 
+ \overline{\partial}^{\mu} \overline{\partial}^{\rho} D(x;x') \; . \qquad 
\label{initialcosmoPi}
\end{eqnarray}
The scalar coefficient functions $A$, $B$ and $D$ are all reflection invariant,
\begin{equation}
A(x;x') = A(x';x) \quad , \quad B(x;x') = B(x';x) \quad , \quad D(x;x') = D(x';x)
\; . \label{ReflectionPi}
\end{equation}
We also remind the reader that the presence of a bar over a tensor indicates 
suppression of its temporal components,
\begin{equation}
\overline{\eta}^{\mu\rho} \equiv \eta^{\mu\rho} + \delta^{\mu}_{~0} 
\delta^{\rho}_{~0} \qquad , \qquad \overline{\partial}^{\mu} \equiv \partial^{\mu}
+ \delta^{\mu}_{~0} \partial_0 \; . \label{bardef}
\end{equation}

Because the vacuum polarization is a bi-vector density, its divergence on each index
group must vanish, on any background geometry. On cosmological backgrounds 
(\ref{geometry}) the divergence produces two independent tensors,
\begin{eqnarray}
\lefteqn{ 0 = \partial_{\mu} \!\times\! i\Bigl[ \mbox{}^{\mu} \Pi^{\rho}_{\rm cos}
\Bigr](x;x') = \overline{\partial}^{\rho} \Bigl[A(x;x') + \partial_0 C(x;x') }
\nonumber \\
& & \hspace{4cm} + \nabla^2 D(x;x')\Bigr] + \delta^{\rho}_{~0} \Bigl[ \partial_0 
B(x;x') \!-\! \nabla^2 C(x';x) \Bigr] \; , \qquad \label{divergencecosmoPi}
\end{eqnarray}
where $\nabla^2 \equiv \partial^i \partial^i$ is the flat space Laplacian.
Expression (\ref{divergencecosmoPi}) allows us to solve for two of the 
coefficient functions,
\begin{equation}
A(x;x') = -\partial_0 C(x;x') - \nabla^2 D(x;x') \qquad , \qquad B(x;x') = 
\frac{\nabla^2}{\partial_0} C(x';x) \; . \label{cosmoPiAB}
\end{equation}
The reflection invariance (\ref{ReflectionPi}) of $A$ and $B$ also implies an
important relation for reflecting $C(x;x')$,
\begin{equation}
\partial_0 C(x;x') = \partial_0' C(x';x) \qquad \Longrightarrow \qquad
C(x';x) = \frac{\partial_0}{\partial_0'} C(x;x') \; . \label{cosmoPiC}
\end{equation}
Note that $\frac1{\partial_0'} C(x;x')$ is reflection invariant.

Substituting (\ref{cosmoPiAB}) and (\ref{cosmoPiC}) into (\ref{initialcosmoPi}) 
shows how the cosmological vacuum polarization can be expressed using two 
structure functions,
\begin{eqnarray}
\lefteqn{ i \Bigl[\mbox{}^{\mu} \Pi^{\rho}_{\rm cos}\Bigr](x;x') = \Bigl[-
\overline{\eta}^{\mu\rho} \partial_0 \partial_0' \!+\! \delta^{\mu}_{~0}
\partial_0' \overline{\partial}^{\rho} \!-\! \overline{\partial}^{\mu} 
\delta^{\rho}_{~0} \partial_0 \!+\! \delta^{\mu}_{~0} \delta^{\rho}_{~0} 
\nabla^2 \Bigr] \frac1{\partial_0'} C(x;x') } \nonumber \\
& & \hspace{7cm} + \Bigl[ \overline{\partial}^{\mu} \overline{\partial}^{\rho}
\!-\! \overline{\eta}^{\mu\rho} \nabla^2\Bigr] D(x;x') \; , \qquad 
\label{cosmoPi1} \\
& & \hspace{2cm} = \Bigl[ \eta^{\mu\nu} \eta^{\rho\sigma} \!-\! \eta^{\mu\sigma}
\eta^{\nu\rho}\Bigr] \partial_{\nu} \partial_{\sigma}' \Bigl[ 
\frac1{\partial_0'} C(x;x')\Bigr] \nonumber \\
& & \hspace{3.5cm} + \Bigl[ \overline{\eta}^{\mu\nu} \overline{\eta}^{\rho\sigma} 
\!-\! \overline{\eta}^{\mu\sigma} \overline{\eta}^{\nu\rho}\Bigr] \partial_{\nu} 
\partial_{\sigma}' \Bigl[D(x;x') \!-\! \frac1{\partial_0'} C(x;x')\Bigr] \; . 
\qquad \label{cosmoPi2}
\end{eqnarray}
The representation (\ref{cosmoPi2}) was first employed in studying charged MMC
scalar contributions to the vacuum polarization on de Sitter background
\cite{Prokopec:2002jn,Prokopec:2002uw}. The procedure for transforming to other 
representations has been given in detail \cite{Leonard:2012si,Leonard:2012ex}.

\section{Graviton Self-Energy in Cosmology}

The purpose of this section is to present our formalism for representing the
graviton self-energy in cosmology. We give a unified derivation which applies
to the cases of matter contributions and to those from gravity itself. We then
specialize the effective field equations to the cases of the graviton mode 
function and to the two potentials that represent the gravitational response
to a point mass.

\subsection{Our Representation}

As discussed in the Introduction, the symmetries of cosmology permit us to 
represent the graviton self-energy as the sum (\ref{initialrep}) of the 21 
tensor differential operators $[\mbox{}^{\mu\nu} \mathcal{D}^{\rho\sigma}_{i}]$ 
of Table~\ref{Tbasis} acting on scalar coefficient functions $T^i(x;x')$,
\begin{equation}
-i\Bigl[\mbox{}^{\mu\nu} \Sigma^{\rho\sigma}\Bigr](x;x') = \sum_{i=1}^{21}
\Bigl[ \mbox{}^{\mu\nu} \mathcal{D}^{\rho\sigma}\Bigr] \!\times\! T^{i}(x;x')
\; . \label{initialrep2}
\end{equation}
Seven of the coefficient functions are related to others by reflection invariance,
as described in Table~\ref{ReflectionT}. Acting the Ward operator (\ref{Wardop})
on a single index group produces a sum (\ref{selfmatter}) of the 10 tensor 
differential operators of Table~\ref{Sbasis} acting on scalar coefficient 
functions $S^i(x;x')$,
\begin{equation}
\mathcal{W}^{\mu}_{~\alpha\beta}(x) \!\times\! -i \Bigl[\mbox{}^{\alpha\beta} 
\Sigma^{\rho\sigma}\Bigr](x;x') = \sum_{i=1}^{10} \Bigl[ \mbox{}^{\mu} 
\mathcal{D}^{\rho\sigma}\Bigr] \!\times\! S^{i}(x;x') \; . \label{selfmatter2}
\end{equation}
Relation (\ref{selfmatter}) can be used to express each of the $S^{i}(x;x')$ in 
terms of the 14 algebraically distinct $T^{j}(x;x')$. The expansions are given 
in Table~\ref{SExpansion}.
\begin{table}[H]
\setlength{\tabcolsep}{8pt}
\def\arraystretch{1.5}
\centering
\begin{tabular}{|@{\hskip 1mm }c@{\hskip 1mm }||c|}
\hline
$S^i(x;x')$ & Expansion in $T^j = T^j(x;x')$ and $T^{jR} = T^j(x';x)$ \\
\hline\hline
$S^1$ & $(D\!-\!1) a H T^3 + (\partial_0 \!-\! a H) T^{13} 
- \frac12 \nabla^2 T^{14R} + a H \nabla^2 T^{16R}$ \\
\hline
$S^2$ & $(\frac{D-1}{2}) a H T^5 + \frac14 T^9 - \frac12 a H T^{10R}$ \\
$$ & $+ \frac12 (\partial_0 - a H) T^{14} + \frac14 \nabla^2 T^{18} 
- \frac12 a H \nabla^2 T^{19R}$ \\
\hline
$S^3$ & $(D\!-\!1) a H T^1 + a H T^2 + (\partial_0 \!-\! a H) T^{3R}
-\frac12 \nabla^2 T^{5R} + a H \nabla^2 T^{7R}$ \\
\hline
$S^4$ & $(D\!-\!1) a H T^{7} + \frac12 T^{10} + a H T^{12}$ \\
$$ & $+ (\partial_0 \!-\! aH) T^{16} + \frac12 \nabla^2 T^{19} + a H \nabla^2 T^{21}$ \\
\hline
$S^5$ & $T^3 - \frac12 \partial_0 T^{14R} + \nabla^2 T^{16R}$ \\
\hline
$S^6$ & $\frac14 \partial_0 T^9 - \frac14 \nabla^2 T^{10R}$ \\
\hline
$S^7$ & $\frac12 T^5 - \frac14 T^{10R} + \frac14 \partial_0 T^{18} 
-\frac12 \nabla^2 T^{19R}$ \\
\hline
$S^8$ & $\frac12 T^2 + \frac14 \partial_0 T^{10} + \frac14 \nabla^2 T^{12}$ \\
\hline
$S^9$ & $T^1 - \frac12 \partial_0 T^{5R} + \nabla^2 T^{7R}$ \\
\hline
$S^{10}$ & $T^7 + \frac12 T^{12} + \frac12 \partial_0 T^{19} + \nabla^2 T^{21}$ \\
\hline
\end{tabular}
\caption{\footnotesize Expansion of the coefficients $S^i(x;x')$ of equation 
(\ref{selfmatter}) in terms of the coefficients $T^j(x;x')$ of the initial
representation (\ref{initialrep}).} 
\label{SExpansion}
\end{table}

Our strategy for representing the $T^{i}(x;x')$ is motivated by the flat space 
analog considered in section \ref{flatgrav}. It is the same for both matter and 
gravity: we use the ten relations of Table~\ref{SExpansion} to solve for the 
coefficient functions in terms of the functions $S^i$ and a ``minimal'' set of 
$T^{i}$'s consisting of $T^{12}$, $T^{16}$, $T^{18}$ and $T^{19}$. Each of the 
ten $S^{i}$ must vanish for matter contributions, whereas they can be nonzero 
for contributions from gravity itself. However, because we must get zero from 
acting the Ward operator on both index groups, and because this action results 
in five distinct tensor operators (\ref{selfgravity}), the ten $S^{i}$ are 
subject to five relations given in Table~\ref{RExpansion}. 
\begin{table}[H]
\setlength{\tabcolsep}{8pt}
\def\arraystretch{1.5}
\centering
\begin{tabular}{|@{\hskip 1mm }c@{\hskip 1mm }||c|}
\hline
$R^i(x;x')$ & Expansion in $S^j = S^j(x;x')$ \\
\hline\hline
$R^1$ & $(\partial_0' \!-\! a' H') S^{1} - \nabla^2 S^{2} + (D\!-\!1) a' H' S^{3}
+ a' H' \nabla^2 S^{4}$ \\
\hline
$R^2$ & $\partial_0' S^{6} - \nabla^2 S^{8}$ \\
\hline
$R^3$ & $\partial_0' S^{2} - S^{3} - \nabla^2 S^{4}$ \\
\hline
$R^4$ & $\!(\partial_0' \!-\! a' H') S^{5} \!-\! S^{6} \!-\! \nabla^2 S^{7} \!+\! 2 a' H' S^{8}
\!+\! (D\!-\!1) a' H' S^{9} \!+\! a' H' \nabla^2 S^{10}\!$ \\
\hline
$R^5$ & $\partial_0' S^{7} - S^{8} - S^{9} - \nabla^2 S^{10}$ \\
\hline
\end{tabular}
\caption{\footnotesize Expansion of the coefficients $R^i(x;x') = 0$ of the Ward
identity (\ref{selfgravity}) in terms of the coefficients $S^j(x;x')$ of the action
(\ref{selfmatter}) of the Ward operator on a single index group.} 
\label{RExpansion}
\end{table}
\noindent We can eliminate $S^3$ using $R^3(x;x') = 0$,
\begin{equation}
S^3 = \partial_0' S^2 - \nabla^2 S^4 \; . \label{S3}
\end{equation}
Combining this with $R^1(x;x') = 0$ gives $S^1$,\footnote{The inverse of $\partial_0'
- a' H'$ can be expressed as a simple integral with respect to $\eta'$,
\begin{equation}
\frac1{\partial_0' \!-\! a' H'} = a' \!\times\! \frac1{\partial_0'} \!\times\! 
\frac1{a'} \; . \label{intseimp}
\end{equation}}
\begin{equation}
S^1 = \frac1{\partial_0' \!-\! a' H'} \Bigl\{ -\Bigl[(D \!-\! 1) a' H' \partial_0'
\!-\! \nabla^2\Bigr] S^2 + (D \!-\! 2) a' H' \nabla^2 S^4\Bigr\} \; . \label{S1}
\end{equation}
The relations $R^2(x;x') = 0$ and $R^5(x;x') = 0$ imply,
\begin{equation}
S^6 = \frac{\nabla^2}{\partial_0'} S^8 \qquad , \qquad S^9 = \partial_0' S^7 - S^8
- \nabla^2 S^{10} \; . \label{S69}
\end{equation}
and substituting (\ref{S69}) in $R^4(x;x') = 0$ gives,
\begin{eqnarray}
\lefteqn{S^5 = \frac1{\partial_0' \!-\! a' H'} \Biggl\{ -\Bigl[(D \!-\!1) a' H' 
\partial_0' \!-\! \nabla^2\Bigr] S^7 } \nonumber \\
& & \hspace{3.5cm} + \Bigl[(D \!-\! 3) a' H' \!+\! \frac{\nabla^2}{\partial_0'}
\Bigr] S^8 + (D\!-\!2) a' H' \nabla^2 S^{10}\Biggr\} \; . \qquad \label{S5}
\end{eqnarray}
So our structure functions consist of $T^{12}$, $T^{16}$, $T^{18}$ and $T^{19}$, plus 
(for graviton contributions) $S^{2}$, $S^{4}$, $S^{7}$, $S^{8}$ and $S^{10}$. 

Because the ten relations in Table~\ref{SExpansion} are coupled they are best solved
in four stages. First use the relations for $S^5$, $S^8$, $S^9$ and $S^{10}$ to write,
\begin{eqnarray}
T^3 & = & \frac12 \partial_0 T^{14R} - \nabla^2 T^{16R} + S^{5} \; , \qquad
\label{T3initial} \\
T^2 & = & -\frac12 \partial_0 T^{10} - \frac12 \nabla^2 T^{12} + 2 S^{8} \; , \qquad
\label{T2initial} \\
T^1 & = & \frac12 \partial_0 T^{5R} - \nabla^2 T^{7R} + S^{9} \; , \qquad 
\label{T1initial} \\
\nabla^2 T^{21} & = & -T^7 -\frac12 T^{12} - \frac12 \partial_0 T^{19} + S^{10} \; .
\qquad \label{T21initial}
\end{eqnarray}
Because $T^{21}$ and $T^{12}$ are symmetric we can use relation \ref{T21initial}) 
to solve for the anti-symmetric part of $T^{7} \equiv T^{7S} + T^{7A}$,
\begin{equation}
T^{7A} \equiv \frac12 (T^{7} - T^{7R}) = -\frac14 (\partial_0 T^{19} - \partial_0' 
T^{19R}) + S^{10A} \; . \label{T7A} 
\end{equation}

The next step is using the $S^2$ and $S^7$ relations to solve for $T^5$ and $T^9$,
\begin{eqnarray}
T^5 & = & \frac12 T^{10R} - \frac12 \partial_0 T^{18} + \nabla^2 T^{19R} + 2 S^7 \; ,
\qquad \label{T5initial} \\
T^9 & = & -(D\!-\!3) a H T^{10R} - 2 (\partial_0 \!-\! a H) T^{14} + 
\Bigl[(D\!-\!1) a H \partial_0 \!-\! \nabla^2\Bigr] T^{18} \nonumber \\
& & \hspace{3cm} - 2(D\!-\!2) a H \nabla^2 T^{19R} + 4 S^2 - 4 (D\!-\!1) a H S^{7}
\; . \qquad \label{T9initial}
\end{eqnarray}
Relation (\ref{T5initial}) could be used in (\ref{T1initial}) to reduce $T^1$ but 
we postpone this. In the 3rd stage the $S^1$ and $S^4$ relations give $T^{13}$ and 
$T^{10}$,
\begin{eqnarray}
T^{13} & = & \frac1{\partial_0 \!-\! a H} \Biggl\{ -\frac12 \Bigl[ (D\!-\!1) a H 
\partial_0 - \nabla^2\Bigr] T^{14R} \nonumber \\
& & \hspace{3cm} + (D\!-\!2) a H \nabla^2 T^{16R} + S^1 - (D\!-\!1) a H 
S^5 \Biggr\} \; , \qquad \label{T13initial} \\
T^{10} & = & -2 (D\!-\! 2) a H T^7 - a H T^{12} \nonumber \\
& & \hspace{2cm} - 2 (\partial_0 \!-\! a H) T^{16} \!+\! (a H \partial_0 \!-\! 
\nabla^2) T^{19} \!+\! 2 S^4 \!-\! 2 a H S^{10} . \qquad \label{T10initial}
\end{eqnarray}
We now use relations (\ref{T5initial}) and (\ref{T10initial}) to update $T^1$, 
$T^2$, $T^5$ and $T^9$,
\begin{eqnarray}
T^{1} & = & -\Bigl[ \Bigl(\frac{D\!-\!2}{2}\Bigr) \partial_0 a H \!+\! \nabla^2
\Bigr] T^7 -\frac14 \partial_0 a H T^{12} -\frac12 \partial_0 (\partial_0 \!-\! a H)
T^{16} \nonumber \\
& & \hspace{1.5cm} - \frac14 \partial_0 \partial_0' T^{18} + \frac14 (\partial_0 a H 
\!-\! \nabla^2) \partial_0 T^{19} + \frac12 \nabla^2 \partial_0' T^{19R} 
\nonumber \\
& & \hspace{1.5cm} + \frac12 \partial_0 S^4 + \partial_0' S^7 + \partial_0 S^{7R} - 
S^8 - \frac12 \partial_0 a H S^{10} - \nabla^2 S^{10R} \; , \qquad \label{T1update} \\
T^{2} & = & (D\!-\!2) \partial_0 a H T^{7} + \frac12 (\partial_0 a H \!-\! \nabla^2)
T^{12} + \partial_0 (\partial_0 \!-\! a H) T^{16} \nonumber \\
& & \hspace{2.6cm} -\frac12 (\partial_0 a H \!-\! \nabla^2) \partial_0 T^{19} 
-\partial_0 S^{4} + 2 S^{8} + \partial_0 a H S^{10} \; , \qquad \label{T2update} \\
T^{5} & = & -(D\!-\!2) a' H' T^{7R} - \frac12 a' H' T^{12} - (\partial_0' \!-\! a' H')
T^{16R} - \frac12 \partial_0 T^{18} \nonumber \\
& & \hspace{2cm} + \frac12 (a' H' \partial_0' \!+\! \nabla^2) T^{19R} + S^{4R} +
2 S^{7} - a' H' S^{10R} \; , \qquad \label{T5update} \\
T^{9} & = & 2(D\!-\!3) (D\!-\!2) a H a' H' T^{7R} + (D\!-\!3) a H a' H' T^{12} -
2 (\partial_0 \!-\! a H) T^{14} \nonumber \\
& & \hspace{1cm} + 2 (D\!-\!3) a H (\partial_0' \!-\! a' H') T^{16R} \!+\! 
\Bigl[(D\!-\!1) a H \partial_0 \!-\! \nabla^2\Bigr] T^{18} \nonumber \\
& & \hspace{1cm} - \Bigl[(D\!-\!3) a H a' H' \partial_0' \!+\! (D\!-\!1) a H 
\nabla^2 \Bigr] T^{19R} + 4 S^{2} \nonumber \\
& & \hspace{1cm} - 2 (D\!-\!3) a H S^{4R} \!-\! 4 (D\!-\!1) a H S^{7} \!+\! 
2 (D\!-\!3) a H a' H' S^{10R} \; . \qquad \label{T9update}
\end{eqnarray}

The final stage begins by noting that the $S^3$ and $S^5$ relations can be 
expressed in terms of two functions $A(x;x')$ and $B(x;x')$,
\begin{eqnarray}
S^3 & \qquad \Longrightarrow \qquad & 0 = -\Bigl[(D\!-\!3) a H \partial_0 \!+\! 
\nabla^2\Bigr] A + \partial_0' B \; , \qquad \label{S3final} \\
S^6 & \qquad \Longrightarrow \qquad & 0 = +\Bigl[(D\!-\!3) \partial_0 a H \!+\! 
\nabla^2\Bigr] A^{R} - \partial_0 B \; . \qquad \label{S6final}
\end{eqnarray}
The functions $A$ and $B$ are,
\begin{eqnarray}
A & \equiv & \frac12 (D\!-\!2) a H T^{7} + \frac14 a H T^{12} + \frac12 (\partial_0 
\!-\! a H) T^{16} -\frac14 (a H \partial_0 \!-\! \nabla^2) T^{19} \nonumber \\
& & \hspace{6.5cm} -\frac12 S^{4} - \frac1{\partial_0} S^{8R} + \frac12a H S^{10} 
\; , \qquad \label{Adef} \\
B & \equiv & \frac12 (\partial_0 \!-\! a H) T^{14} - \frac14 \Bigl[ (D\!-\!1) a H 
\partial_0 \!-\! \nabla^2\Bigr] T^{18} + \frac{(D\!-\!2)}{2} a H \nabla^2 T^{19R}
\nonumber \\
& & \hspace{4.4cm} - S^{2} + (D\!-\! 1) a H S^{7} - (D\!-\!3) a H \frac1{\partial_0'}
S^{8} \; . \qquad \label{Bdef}
\end{eqnarray}
We first solve equation (\ref{S3final}) for $T^{14}$,
\begin{eqnarray}
\lefteqn{ T^{14} = \frac1{\partial_0 \!-\! a H} \Biggl\{ \frac12 \Bigl[(D\!-\!1) a H
\partial_0 \!-\! \nabla^2\Bigr] T^{18} - (D\!-\!2) a H \nabla^2 T^{19R} + 2 S^{2} }
\nonumber \\
& & \hspace{0.5cm} -2 (D\!-\! 1) a H S^{7} + 2 (D\!-\!3) a H \frac1{\partial_0'} S^{8}
+ \Bigl[ (D\!-\!3) a H \partial_0 \!+\! \nabla^2\Bigr] \frac1{\partial_0'} A 
\Biggr\} . \qquad \label{T14}
\end{eqnarray}
The final relation derives from combining (\ref{S3final}) with (\ref{S6final}),
\begin{equation}
\partial_0 \Bigl( {\rm Eqn.}~\ref{S3final}\Bigr) - \partial_0' \Bigl( {\rm 
Eqn.}~\ref{S6final}\Bigr) = \Bigl[ (D\!-\!3) \partial_0 a H \!+\! \nabla^2\Bigr]
\Bigl[-\partial_0 A + \partial_0' A^R\Bigr] = 0 \; . \label{T7Sinitial}
\end{equation}
It follows that $\partial_0 A = \partial_0' A^R$, which allows us to solve for the
symmetric part of $T^{7}$. Combining this with the antisymmetric part (\ref{T7A}) 
gives,
\begin{equation}
T^{7} = - \frac{\frac12 T^{12}}{D\!-\!2} - \frac12 \partial_0 T^{19} + S^{10} - 
\frac{2}{D \!-\! 2} \frac1{(\partial_0 a H \!-\! \partial_0' a' H')} \Bigl[ 
\partial_0 \Delta A \!-\! \partial_0' \Delta A^{R}\Bigr] \; . \label{T7final}
\end{equation}
Here the residual part of $A$ is,
\begin{eqnarray}
\lefteqn{ \Delta A \equiv \frac12 (\partial_0 \!-\! a H) T^{16} - \frac14 \Bigl[ 
(D\!-\! 1) a H \partial_0 \!-\! \nabla^2\Bigr] T^{19} } \nonumber \\
& & \hspace{5.9cm} - \frac12 S^4 - \frac1{\partial_0} S^{8R} + \frac{(D\!-\!1)}{2} a H
S^{10} \; . \qquad \label{DeltaA}
\end{eqnarray}

We should comment on how to invert the differential operator $\mathcal{D} =
\partial_0 a H - \partial_0' a' H'$. This is accomplished by first factoring out 
$a H \times a' H'$,
\begin{equation}
\mathcal{D} = \frac1{a H \, a' H'} \times \frac1{a H \partial_0 \!-\! a' H' 
\partial_0'} \times a H \, a' H' \; . \label{hardD}
\end{equation}
Now change the time variable from $\eta$ to $u$ such that,
\begin{equation}
du \equiv \frac{d\eta}{a H} \; . \label{etatou}
\end{equation}
By employing ``lightcone'' variables,
\begin{equation}
u_{\pm} \equiv \frac12 (u \pm u') \qquad \Longrightarrow \qquad 
\frac{\partial}{\partial u} - \frac{\partial}{\partial u'} = 
\frac{\partial}{\partial u_{-}} \; , \label{lightcone}
\end{equation}
we can express (\ref{hardD}) as an integration with respect to $u_{-}$,
\begin{equation}
\frac1{\partial_0 a H \!-\! \partial_0'a' H'} = \frac1{a H \, a' H'} \times
\int \!\! du_{-} \times a H \, a' H' \; . \label{simpleD}
\end{equation}
For the important special case of de Sitter we have $u = -\frac12 \eta^2$.

Our final expressions for the coefficient functions can be simplified by using 
two symmetric auxiliary functions to absorb all the terms involving the inverse 
of $\mathcal{D}$. We first define the (not necessarily symmetric) function 
$\gamma(x;x')$,
\begin{eqnarray}
\lefteqn{\gamma \equiv \partial_0 (\partial_0 \!-\! a H) T^{16} - \frac12 
\Bigl[(D \!-\! 1) \partial_0 a H \!-\! \nabla^2\Bigr] \partial_0 T^{19} }
\nonumber \\
& & \hspace{6cm} - \partial_0 S^4 - 2 S^{8R} + (D\!-\!1) \partial_0 a H S^{10}
\; . \qquad \label{gammadef}
\end{eqnarray}
The two symmetric functions are, 
\begin{eqnarray}
\alpha & \equiv & \frac1{\mathcal{D}} \Bigl[ \gamma - \gamma^{R}\Bigr] \; , 
\qquad \label{alphadef} \\
\beta & \equiv & \frac12 \Bigl(\gamma \!+\! \gamma^{R}\Bigr) - \frac12 \Bigl(
\partial_0 a H \!+\! \partial_0' a' H'\Bigr) \alpha \; . \qquad \label{betadef}
\end{eqnarray}
Note that the function $\beta(x;x')$ can be written in two different ways,
\begin{equation}
\beta = \gamma - \partial_0 a H \alpha = \gamma^{R} - \partial_0' a' H' \alpha
\; . \label{alphadif}
\end{equation}
Also note that we can eliminate $T^{16}$,
\begin{eqnarray}
\lefteqn{T^{16} = \frac1{\partial_0 \!-\! a H} \frac1{\partial_0} \Biggl\{ 
\beta + \partial_0 a H \alpha + \frac12 \Bigl[ (D \!-\! 1) \partial_0 a H \!-\!
\nabla^2 \Bigr] \partial_0 T^{19} } \nonumber \\
& & \hspace{5.5cm} + \partial_0 S^4 + 2 S^{8R} - (D\!-\! 1) \partial_0 a H S^{10}
\Biggr\} . \qquad \label{T16final}
\end{eqnarray} 

The notation can be further simplified by introducing symbols to stand for three
differential operators and an inverse operator that occur repeatedly,
\begin{eqnarray}
D_0 \equiv \partial_0 a H \qquad & , & \qquad D_1 \equiv (D\!-\!1) \partial_0 a H
- \nabla^2 \; , \label{DOPS1} \\
\mathcal{I} \equiv \frac1{\partial_0 \!-\! a H} \frac1{\partial_0} \qquad & , &
\qquad D_3 \equiv (D \!-\!3) \partial_0 a H + \nabla^2 \; . \label{DOPS2}
\end{eqnarray}
Giving any of these operators a prime indicates that it is constructed from the
same quantities at ${x'}^{\mu}$ instead of $x^{\mu}$, for example, $D_0' =
\partial_0' a' H'$. With these definitions our final expressions for the 
algebraically independent coefficient functions are,
\begin{eqnarray}
\lefteqn{T^1 = \frac{\nabla^2 T^{12}}{2 (D\!-\! 2)} - \frac14 \partial_0 
\partial_0' T^{18} + \frac12 \nabla^2 \Bigl(\partial_0 T^{19} \!+\! \partial_0'
T^{19R} \Bigr) + \frac{\nabla^2 \alpha}{D \!-\!2} - \frac12 \beta } \nonumber \\
& & \hspace{2.9cm} + \Bigl( \partial_0' S^{7} \!+\! \partial_0 S^{7R}\Bigr) -
\Bigl( S^{8} \!+\! S^{8R}\Bigr) - \nabla^2 \Bigl( S^{10} \!+\! S^{10R} \Bigr) 
\; , \qquad \label{T1final} \\
\lefteqn{T^2 = -\frac12 \nabla^2 T^{12} + \beta + 2 \Bigl( S^{8} \!+\! S^{8R}
\Bigr) \; ,} \label{T2final} \\
\lefteqn{T^3 = \mathcal{I}' \Biggl\{ \frac14 D_1' \partial_0 \partial_0' T^{18} 
- \frac12 (D\!-\!2) \nabla^2 D_0' \partial_0 T^{19} - \frac12 \nabla^2 D_1' 
\partial_0' T^{19R} - \nabla^2 D_0' \alpha } \nonumber \\
& & \hspace{0cm} + \Bigl(\frac12 D_1' \!-\! D_0'\Bigr) \beta + \partial_0 
\partial_0' S^{2R} - \nabla^2 \partial_0' S^{4R} - D_1' \partial_0' S^7 - 
(D\!-\!1) D_0' \partial_0 S^{7R} \nonumber \\
& & \hspace{0.4cm} + D_3' S^8 + (D\!-\!3) D_0' S^{8R} + (D\!-\! 2) \nabla^2 D_0' 
S^{10} + (D\!-\!1) \nabla^2 D_0' S^{10R} \Biggr\} , \qquad \label{T3final} \\
\lefteqn{T^5 = -\frac12 \partial_0 T^{18} + \nabla^2 T^{19R} - 
\frac1{\partial_0'} \beta + 2 S^7 - \frac{2}{\partial_0'} S^8 \; ,}  
\label{T5final} \\
\lefteqn{T^7 = -\frac{T^{12}}{2 (D\!-\!2)} - \frac12 \partial_0 T^{19} -
\frac{\alpha}{D \!-\! 2} + S^{10} \; , } \label{T7Final} \\ 
\lefteqn{T^9 = -\frac{2 \nabla^2}{\partial_0 \partial_0'} \beta \; , }
\label{T9final} \\
\lefteqn{T^{10} = -\frac{2}{\partial_0} \beta - \frac4{\partial_0} S^{8R} 
\; , } \label{T10final} \\
\lefteqn{T^{13} = \mathcal{I} \mathcal{I}' \Biggl\{ -\frac14 D_1 D_1' \partial_0
\partial_0' T^{18} + \frac12 (D\!-\!2) \nabla^2 \Bigl[D_1 D_0' \partial_0 T^{19} 
\!+\! D_0 D_1' \partial_0' T^{19R} \Bigr] } \nonumber \\
& & \hspace{-0.5cm} + (D\!-\!2) \nabla^2 D_0 D_0' \alpha \!-\! \frac12 \Bigl[ 
(D\!-\!3) D_1 D_3' \!-\! D_3 \nabla^2\Bigr] \beta \!-\! D_1' \partial_0 
\partial_0' S^2 \!-\! D_1 \partial_0 \partial_0' S^{2R} \nonumber \\
& & \hspace{-0.5cm} + (D\!-\!2) \nabla^2 \Bigl[D_0' \partial_0 S^4 \!+\! D_0 
\partial_0' S^{4R}\Bigr] + (D\!-\!1) \Bigl[ D_0 D_1' \partial_0' S^7 \!+\! D_0' 
D_1 \partial_0 S^{7R} \Bigr] \nonumber \\
& & \hspace{-0.5cm} - (D\!-\!3) \Bigl[ D_0 D_1' S^8 \!+\! D_0 D_1 S^{8R}\Bigr]
\!-\! (D\!-\!2) (D\!-\!1) \nabla^2 D_0 D_0' (S^{10} \!+\! S^{10R}) \Biggr\}, 
\qquad \label{T13final} \\
\lefteqn{T^{14} = \mathcal{I} \Biggl\{ \frac12 D_1 \partial_0 T^{18} - (D\!-\!2)
\nabla^2 D_0 T^{19R} + \frac{D_3}{\partial_0'} \beta } \nonumber \\
& & \hspace{4.3cm} + 2 \partial_0 S^2 - 2 (D\!-\!1) D_0 S^7 + 2 (D\!-\!3)
\frac{D_0}{\partial_0'} S^8 \Biggr\} , \qquad \label{T14final} \\
\lefteqn{T^{21} = -\frac12 \Bigl(\frac{D\!-\!3}{D\!-\!2}\Bigr) \frac1{\nabla^2}
T^{12} + \frac1{D\!-\!2} \frac1{\nabla^2} \alpha \; . } \label{T21final}
\end{eqnarray}
The cumbersome nature of these expressions prompts several comments on the
issue of accuracy. First, the flat space limit agrees with the decomposition
of section 2.2. Second, our results for $T^1(x;x')$, $T^2(x;x')$, $T^9(x;x')$
and $T^{13}(x;x')$ are reflection symmetric as they should be. Finally, the
contributions proportional to $T^{12}$, $T^{18}$, $T^{19}$, $T^{19R}$, 
$\alpha$, and $\beta$ are each separately annihilated by the action of the
Ward operator (\ref{Wardop}) on either index group, while the contributions 
proportional to $S^2$, $S^4$, $S^7$, $S^8$ and $S^{10}$ are annihilated when
the Ward operator is acted on both index groups.

\subsection{Ricci and Weyl Operators}

Our results (\ref{T16final}) and (\ref{T1final}-\ref{T21final}) can be 
reorganized into a sum of products of tensor differential operators acting on 
the fundamental structure functions. Each of these tensor differential operators
is separately annihilated by the action of either one or two Weyl operators.
All the operators descend from $3+1$ decomposing the same products of the
transverse projection operator that we encountered in section 2.2,
\begin{equation}
\Pi^{\mu\nu} \equiv \partial^{\mu} \partial^{\nu} - \partial^2 \eta^{\mu\nu} 
= \Pi_{A}^{\mu\nu} + \Pi_{B}^{\mu\nu} \; , \label{PiAB} 
\end{equation}
where the two projectors are,
\begin{eqnarray}
\Pi_{A}^{\mu\nu} & \equiv & \overline{\partial}^{\mu} \overline{\partial}^{\nu}
- \nabla^2 \overline{\eta}^{\mu\nu} \; , \label{PiAdef} \\
\Pi_{B}^{\mu\nu} & \equiv & \nabla^2 \delta^{\mu}_{~0} \delta^{\nu}_{~0} - 2
\partial_0 \delta^{(\mu}_{~~0} \overline{\partial}^{\nu )} + \partial_0^2 
\overline{\eta}^{\mu\nu} \; . \label{PiBdef}
\end{eqnarray}
What we term the {\it Ricci operators} are simple extensions of (\ref{PiAdef})
and (\ref{PiBdef}),
\begin{eqnarray}
\Pi_{A}^{\mu\nu} \longrightarrow \overline{\partial}^{\mu} 
\overline{\partial}^{\nu} - \nabla^2 \overline{\eta}^{\mu\nu} + 
\frac{(D \!-\!2) \nabla^2}{\partial_0 \!-\! a H} a H \delta^{\mu}_{~0}
\delta^{\nu}_{~0} \equiv \mathcal{R}_{A}^{\mu\nu}(x) \; ,
\label{RAdef} \\
\Pi_{B}^{\mu\nu} \longrightarrow \frac{\delta^{\mu}_{~0} \delta^{\nu}_{~0}}{
\partial_0 \!-\! a H} \Bigl[ \nabla^2 \!-\! (D\!-\!1) a H \partial_0\Bigr] 
\partial_0 \!-\! 2 \partial_0 \delta^{(\mu}_{~~0} \overline{\partial}^{\nu )} 
\!+\! \partial_0^2 \overline{\eta}^{\mu\nu} \equiv \mathcal{R}_{B}^{\mu\nu}(x) 
\; . \label{RBdef}
\end{eqnarray}
It is straightforward to verify that acting the Ward operator annihilates 
each Ricci operator,
\begin{equation}
\mathcal{W}^{\mu}_{~\alpha\beta}(x) \!\times\! \mathcal{R}_{A}^{\alpha\beta}(x) 
= 0 = \mathcal{W}^{\mu}_{~\alpha\beta}(x) \!\times\! 
\mathcal{R}_{B}^{\alpha\beta}(x) \; .
\end{equation}

The related {\it Weyl operators} come from extending products of two projection
operators. The simplest is purely spatial,
\begin{equation}
\Pi_{A}^{\mu (\rho} \!\times\! \Pi_{A}^{\sigma )\nu} \longrightarrow 
\Pi_{A}^{\mu (\rho} \Pi_{A}^{\sigma )\nu} - \frac{\Pi_{A}^{\mu\nu} 
\Pi_{A}^{\rho\sigma}}{D \!-\! 2} \equiv \mathcal{C}_{AA}^{\mu\nu\rho\sigma} \; .
\label{CAAdef}
\end{equation}
Because $\mathcal{C}_{AA}^{\mu\nu\rho\sigma}$ is both transverse and traceless,
it is annihilated when the Ward operator acts on either index group,
\begin{equation}
\mathcal{W}^{\mu}_{~\alpha\beta}(x) \!\times\! \mathcal{C}_{AA}^{\alpha\beta
\rho\sigma} = 0 = \mathcal{W}^{\rho}_{~\gamma\delta}(x') \!\times\! 
\mathcal{C}_{AA}^{\mu\nu\gamma\delta} \; .
\end{equation}
The second Weyl operator comes from extending the product of two $B$-type 
projectors,
\begin{eqnarray}
\lefteqn{\Pi_{B}^{\mu (\rho} \!\times\! \Pi_{B}^{\sigma )\nu} \longrightarrow
\mathcal{C}_{BB}^{\mu\nu\rho\sigma}(x;x') \equiv \partial_0^2 {\partial_0'}^2
\overline{\eta}^{\mu (\rho} \overline{\eta}^{\sigma )\nu} - 2 \partial_0 
{\partial_0'}^2 \delta^{(\mu}_{~~0} \overline{\eta}^{\nu ) (\rho} 
\overline{\partial}^{\sigma )} } \nonumber \\
& & \hspace{0cm} + 2 \partial_0^2 \partial_0' \overline{\partial}^{(\mu}
\overline{\eta}^{\nu) (\rho} \delta^{\sigma)}_{~~0} - 2 \partial_0 \partial_0'
\nabla^2 \delta^{(\mu}_{~~0} \overline{\eta}^{\nu)(\rho} \delta^{\sigma)}_{~~0} 
- 2 \partial_0 \partial_0' \delta^{(\mu}_{~~0} \overline{\partial}^{\nu)}
\delta^{(\rho}_{~~0} \overline{\partial}^{\sigma)} \nonumber \\
& & \hspace{-0.7cm} - \frac{\delta^{\mu}_{~0} \delta^{\nu}_{~0}}{\partial_0 \!-\!
a H} a H \partial_0^2 \Bigl[{\partial_0'}^2 \overline{\eta}^{\rho\sigma} \!\!+\!
2 \partial_0' \delta^{(\rho}_{~~0} \overline{\partial}^{\sigma)} \Bigr] -
\Bigl[\partial_0^2 \overline{\eta}^{\mu\nu} \!\!-\! 2 \partial_0 \delta^{(\mu}_{~~0}
\overline{\partial}^{\nu)}\Bigr] \frac{ \delta^{\rho}_{~0} \delta^{\sigma}_{~0}
}{\partial_0' \!-\! a' H'} a' H' {\partial_0'}^2 \nonumber \\ 
& & \hspace{-0.7cm} + \frac{\delta^{\mu}_{~0} \delta^{\nu}_{~0}}{\partial_0 \!-\!
a H} \partial_0 \Bigl[\partial_0' \overline{\partial}^{\rho} 
\overline{\partial}^{\sigma} \!\!+\! 2 \nabla^2 \delta^{(\rho}_{~~0} 
\overline{\partial}^{\sigma)} \Bigr] \partial_0' + \Bigl[\partial_0 
\overline{\partial}^{\mu} \overline{\partial}^{\nu} \!\!-\! 2 \nabla^2 
\delta^{(\mu}_{~~0} \overline{\partial}^{\nu)}\Bigr] \partial_0 
\frac{ \delta^{\rho}_{~0} \delta^{\sigma}_{~0}}{\partial_0' \!-\! a' H'} 
\partial_0' \nonumber \\
& & \hspace{-0.7cm} + \frac{\delta^{\mu}_{~0} 
\delta^{\nu}_{~0} \delta^{\rho}_{~0} \delta^{\sigma}_{~0}}{(\partial_0 \!-\! a H) 
(\partial_0' \!-\! a' H')} \! \Bigl[(D\!-\!1) a H \partial_0 a' \!H' \partial_0' 
\!\!-\! (a H \partial_0 \!+\! a' H' \partial_0') \nabla^2 \!\!+\!\! \nabla^4\!
\Bigr] \partial_0 \partial_0' . \quad \label{CBBdef} 
\end{eqnarray}
One can also show that $\mathcal{C}_{BB}^{\mu\nu\rho\sigma}$ is annihilated by
the Ward operator acting on either index group,
\begin{equation}
\mathcal{W}^{\mu}_{~\alpha\beta}(x) \!\times\! \mathcal{C}_{BB}^{\alpha\beta
\rho\sigma}(x;x') = 0 = \mathcal{W}^{\rho}_{~\gamma\delta}(x') \!\times\! 
\mathcal{C}_{BB}^{\mu\nu\gamma\delta}(x;x') \; .
\end{equation}

We can use the Ricci and Weyl operators to express matter contributions to the
graviton self-energy,
\begin{eqnarray}
\lefteqn{ -i \Bigl[ \mbox{}^{\mu\nu} \Sigma^{\rho\sigma}_{\rm mat}\Bigr](x;x')
= -\frac12 \mathcal{C}_{AA}^{\mu\nu\rho\sigma} \!\times\! \frac1{\nabla^2} T^{12} 
- \frac14 \mathcal{R}_{B}^{\mu\nu}(x) \!\times\! \mathcal{R}_{B}^{\rho\sigma}(x') 
\!\times\! \frac1{\partial_0 \partial_0'} T^{18} } \nonumber \\
& & \hspace{0cm} + \frac12 \Bigl[ \mathcal{R}_{B}^{\mu\nu}(x) \!\times\! 
\mathcal{R}_{A}^{\rho\sigma}(x') \!\times\! \frac1{\partial_0} T^{19} + 
\mathcal{R}_{A}^{\mu\nu}(x) \!\times\! \mathcal{R}_{B}^{\rho\sigma}(x') \!\times\!
\frac1{\partial_0'} T^{19R} \Bigr] \nonumber \\
& & \hspace{-0.5cm} + \frac{\mathcal{R}_{A}^{\mu\nu}(x) \!\times\! 
\mathcal{R}_{A}^{\rho\sigma}(x')}{D\!-\!2} \frac1{\nabla^2} \alpha + \Bigl[ 
\mathcal{C}_{BB}^{\mu\nu\rho\sigma}(x;x') \!-\! \frac12 \mathcal{R}_{B}^{\mu\nu}(x) 
\!\times\! \mathcal{R}_{B}^{\rho\sigma}(x') \Bigr] \frac1{\partial_0^2 
{\partial_0'}^2} \beta \; . \qquad \label{mattersigma}
\end{eqnarray}
Expressing contributions from gravity itself requires the additional operators
formed from suppressing the temporal components of $\mathcal{C}_{BB}^{\mu\nu\rho
\sigma}(x;x')$ on either $x^{\mu}$ of ${x'}^{\mu}$,
\begin{equation}
\mathcal{C}_{BB}^{\overline{\mu} \overline{\nu} \rho\sigma}(x;x') \equiv
\frac{\overline{\delta}^{\mu}_{~\alpha} \overline{\delta}^{\nu}_{~\beta}}{
\partial_0^2} \mathcal{C}_{BB}^{\alpha\beta\rho\sigma}(x;x') \;\; , \;\; 
\mathcal{C}_{BB}^{\mu\nu \overline{\rho} \overline{\sigma}}(x;x') \equiv
\frac{\overline{\delta}^{\rho}_{~\gamma} \overline{\delta}^{\sigma}_{~\delta}}{
{\partial_0'}^2} \mathcal{C}_{BB}^{\mu\nu\gamma\delta}(x;x') \; ,
\end{equation}
where $\overline{\delta}^{\alpha}_{~\beta} \equiv \delta^{\alpha}_{~\beta}
- \delta^{\alpha}_{~0} \delta^{0}_{~\beta}$ is the spatial unit matrix.
The contributions from gravity itself include all of the same terms in 
(\ref{mattersigma}) with the addition of terms involving the structure functions
$S^2$, $S^4$, $S^7$, $S^8$ and $S^{10}$,
\begin{eqnarray}
\lefteqn{ -i \Bigl[ \mbox{}^{\mu\nu} \Sigma^{\rho\sigma}_{\rm grav}\Bigr](x;x') =
\Bigl({\rm Eqn.}~\ref{mattersigma}\Bigr) } \nonumber \\
& & \hspace{-0.7cm} + \frac{\delta^{\mu}_{~0} \delta^{\nu}_{~0}}{\partial_0 \!-\! 
a H} \Bigl[\mathcal{R}_{B}^{\rho\sigma}(x') \frac1{\partial_0'} S^2 \!\!+\! 
\mathcal{R}_{A}^{\rho\sigma}(x') S^4\Bigr] \!+\! \frac{\delta^{\rho}_{~0} 
\delta^{\sigma}_{~0}}{\partial_0' \!-\! a' H'} \Bigl[ \mathcal{R}_{B}^{\mu\nu}(x) 
\frac1{\partial_0} S^{2R} \!\!+\! \mathcal{R}_{A}^{\mu\nu}(x) S^{4R}\Bigr] 
\nonumber \\
& & \hspace{0cm} + \Bigl[ \overline{\eta}^{\mu\nu} \!-\! \frac{(D \!-\!1) 
\delta^{\mu}_{~0} \delta^{\nu}_{~0}}{\partial_0 \!-\! aH} a H \Bigr] \Bigl[
\mathcal{R}_{B}^{\rho\sigma}(x') \frac1{\partial_0'} S^7 + 
\mathcal{R}_{A}^{\rho\sigma}(x') S^{10} \Bigr] \nonumber \\
& & \hspace{0.5cm} + \Bigl[ \overline{\eta}^{\rho\sigma} \!-\! \frac{(D \!-\!1) 
\delta^{\rho}_{~0} \delta^{\sigma}_{~0}}{\partial_0' \!-\! a' H'} a' H' \Bigr] 
\Bigl[\mathcal{R}_{B}^{\mu\nu}(x) \frac1{\partial_0} S^{7R} + 
\mathcal{R}_{A}^{\mu\nu}(x) S^{10R} \Bigr] \nonumber \\
& & \hspace{1cm} + \Biggl\{ \mathcal{C}_{BB}^{\overline{\mu} \overline{\nu} 
\rho\sigma}(x;x') - \Bigl[ \overline{\eta}^{\mu\nu} - \frac{(D\!-\!3) 
\delta^{\mu}_{~0} \delta^{\nu}_{~0}}{\partial_0 \!-\! a H} a H\Bigr] 
\mathcal{R}_{B}^{\rho\sigma}(x') \Biggr\} \frac1{{\partial_0'}^2} S^8
\nonumber \\
& & \hspace{1.5cm} + \Biggl\{ \mathcal{C}_{BB}^{\mu\nu \overline{\rho} 
\overline{\sigma}}(x;x') - \mathcal{R}_{B}^{\mu\nu}(x) \Bigl[ 
\overline{\eta}^{\rho\sigma} - \frac{(D\!-\!3) \delta^{\rho}_{~0} 
\delta^{\sigma}_{~0}}{\partial_0' \!-\! a' H'} a' H'\Bigr] \Biggr\} 
\frac1{\partial_0^2} S^{8R} . \qquad \label{gravitonsigma}
\end{eqnarray}

\subsection{Effective Field Equations}

Of course the point of developing this representation for $-i [\mbox{}^{\mu\nu} 
\Sigma^{\rho\sigma}](x;x')$ is to facilitate solving the effective field equation 
(\ref{Einsteineqn}). We here adapt it to two important special cases:
\begin{enumerate}
\item{Plane wave gravitons; and}
\item{The response to a point mass.}
\end{enumerate}
In the first case the graviton field $h_{\mu\nu}(x)$ takes the form,
\begin{equation}
{\rm Gravitons} \Longrightarrow h_{\mu\nu} = u(\eta,k) e^{i \vec{k} \cdot \vec{x}} 
\epsilon_{\mu\nu}(\vec{k}) \quad , \quad \epsilon_{0\mu} = 0 = k_i \epsilon_{i\mu} =
\epsilon_{ii} \; , \label{gravitons}
\end{equation}
and what we seek is an equation for the mode function $u(\eta,k)$. For the second 
the graviton field takes the form,
\begin{equation}
{\rm Potentials} \Longrightarrow \kappa h_{00} = -2 \Psi(\eta,r) \; , \; \kappa 
h_{0i} = 0 \; , \; \kappa h_{ij} = -2 \Phi(\eta,r) \delta_{ij} \; , 
\label{potentials}
\end{equation}
and what we seek are equations for the two potentials $\Psi(\eta,r)$ and 
$\Phi(\eta,r)$.

\subsubsection{The Graviton Mode Function}

On a general cosmological background the Lichnerowicz operator receives 
contributions from whatever matter source supports the geometry. On de Sitter
background its action on a general graviton field takes the form
\cite{Park:2015kua},
\begin{equation}
\mathcal{L}^{\mu\nu\rho\sigma} h_{\rho\sigma} = \partial_{\alpha} \Bigl[a^2 
\mathcal{L}^{\mu\nu\rho\sigma\alpha\beta} \partial_{\beta} h_{\rho\sigma}\Bigr]
+ \partial_{\alpha} \Bigl[H a^3 \eta^{\mu\nu} h^{\alpha 0}\Bigr] - H a^3 
\eta^{0 (\mu} \partial^{\nu )} h \; , \label{Lich1}
\end{equation}
where the tensor factor is,
\begin{equation}
\mathcal{L}^{\mu\nu\rho\sigma\alpha\beta} = \frac12 \eta^{\alpha\beta} (
\eta^{\mu (\rho} \eta^{\sigma ) \nu} \!-\! \eta^{\mu\nu} \eta^{\rho\sigma} )
\!+\! \frac12 \eta^{\mu\nu} \eta^{\rho (\alpha} \eta^{\beta )\sigma} \!+\! 
\frac12 \eta^{\rho\sigma} \eta^{\mu (\alpha} \eta^{\beta ) \nu} \!-\! 
\eta^{\alpha (\rho} \eta^{\sigma ) (\mu} \eta^{\nu )\beta} . \label{Lich2}
\end{equation}
Acting the Lichnerowicz operator on a plane wave graviton (\ref{gravitons})
gives,
\begin{equation}
\mathcal{L}^{\mu\nu\rho\sigma} u(\eta,k) e^{i \vec{k} \cdot \vec{x}} 
\epsilon_{\rho\sigma} = -\frac12 a^2 \Bigl[ \partial_0^2 + 2 a H \partial_0
+ k^2\Bigr] u(\eta,k) \times e^{i \vec{k} \cdot \vec{x}} \epsilon^{\mu\nu}
\; . \label{classgraviton}
\end{equation}

Gravitons have zero stress tensor. Because their polarization tensor 
$\epsilon_{\mu\nu}$ is purely spatial, transverse and also traceless, the
only one of the coefficient functions $T^i(x;x')$ that contributes is
$T^2(x;x')$,\footnote{Of course $T^2(x;x')$ may
vanish for some theories, however, it is generally nonzero because gravity
couples to all fields and the 0-point fluctuations of these fields can make
nonzero contributions to $-i[\mbox{}^{\mu\nu} \Sigma^{\rho\sigma}](x;x')$
even if the stress-energy of the background vanishes.}
\begin{equation}
\int \!\! d^4x' \Bigl[ \mbox{}^{\mu\nu} \Sigma^{\rho\sigma}\Bigr](x;x')
h_{\rho\sigma}(x') = \epsilon^{\mu\nu}(\vec{k}) e^{i \vec{k} \cdot \vec{x}}
\int \!\! d^4x' \, i T^2(x;x') u(\eta',k) e^{-i \vec{k} \cdot \Delta \vec{x}} 
\; . \label{gravitonsource}
\end{equation}
Hence the equation for corrections to the graviton mode function is,
\begin{equation}
-\frac12 a^2 \Bigl[ \partial_0^2 + 2 a H \partial_0 + k^2\Bigr] u(\eta,k)
= \int \!\! d^4x' \, i T^2(x;x') u(\eta',k) e^{-i \vec{k} \cdot \Delta \vec{x}} 
\; . \label{modeeqn}
\end{equation}
This is the same for matter and gravity contributions, however, what $T^2$
is in terms for the fundamental structure functions differs according to
relation (\ref{T2final}). Note that it will be necessary to extract a 
number of derivatives from the structure functions; primed derivatives being
acted on the mode function and unprimed derivatives pulled outside the 
integration.

\subsubsection{Response to A Point Mass}

Acting the Lichnerowicz operator on the potentials (\ref{potentials}) produces,
\begin{eqnarray}
\lefteqn{\mathcal{L}^{\mu\nu\rho\sigma} \Bigl[-2 \delta^0_{\rho} 
\delta^0_{\sigma} \Psi -2 \overline{\eta}_{\rho\sigma} \Phi\Bigr] = a^2 
\delta^{\mu}_0 \delta^{\nu}_0 \Bigl[ 6 a^2 H^2 \Psi - 2 (\nabla^2 \!-\! 
3 a H \partial_0) \Phi\Bigr] } \nonumber \\
& & \hspace{0.5cm} + 2 a^2 \delta^{(\mu}_0 \overline{\partial}^{\nu )} \Bigl[
2 a H \Psi + 2 \partial_0 \Phi\Bigr] + a^2 \overline{\partial}^{\mu}
\overline{\partial}^{\nu} \Bigl[ \Psi - \Phi\Bigr] \nonumber \\
& & \hspace{1.5cm} + a^2 \overline{\eta}^{\mu\nu} \Bigl[ - (\nabla^2 \!+\! 2 
a H \partial_0 \!+\! 6 a^2 H^2) \Psi + (\nabla^2 \!-\! 4 a H \partial_0 \!-\!
2 \partial_0^2) \Phi\Bigr] \; . \qquad \label{classpotentials}
\end{eqnarray}
The potentials $\Psi(\eta,r)$ and $\Phi(\eta,r)$ are the response to a static
point mass $M$ whose linearized stress tensor is,
\begin{equation}
\frac12 \kappa T^{\mu\nu}_{\rm lin}(\eta,\vec{x}) = -\frac12 \kappa M a(\eta) 
\delta^3(\vec{x}) \delta^{\mu}_0 \delta^{\nu}_0 \; . \label{pointmass}
\end{equation}
The zeroth order response is,
\begin{equation}
\Psi_0(\eta,r) = \Phi_0(\eta,r) = -\frac{G M}{a(\eta) r} \; . \label{zerothpots}
\end{equation}
Loop corrections are sourced by the integral of the self-energy against
lower order response,
\begin{equation}
\mathcal{L}^{\mu\nu\rho\sigma} h_{\rho\sigma}(x) = \int \!\! d^4x' 
\Bigl[\mbox{}^{\mu\nu} \Sigma^{\rho\sigma}\Bigr](x;x') h_{\rho\sigma}(x')
\equiv \mathcal{S}^{\mu\nu}(x) \; . \label{sourcepots}
\end{equation}
Although we have worked out all components of $\mathcal{S}^{\mu\nu}(x)$, the 
only necessary ones are $\mathcal{S}^{0i}$ --- which is $\partial_i$ of something 
--- and the $\partial_i \partial_j$ part of $\mathcal{S}^{ij}$. Comparison with 
(\ref{classpotentials}) implies that the two potentials obey,
\begin{eqnarray}
\lefteqn{4 a^3 H \Psi + 4 a^2 \partial_0 \Phi = 2 \! \int \!\! d^4x' \Biggl\{
i T^{14}(x';x) \Psi(x') } \nonumber \\
& & \hspace{3cm} + \Bigl[3 i T^5(x';x) \!-\! i T^{10}(x;x') \!-\! 
i T^{19}(x;x') \nabla^2 \Bigr] \Phi(x') \Biggr\} , \qquad \label{onepotA} \\
\lefteqn{a^2 \Psi - a^2 \Phi = -2 \! \int \!\! d^4x' \Biggl\{i T^{16}(x';x) 
\Psi(x') } \nonumber \\
& & \hspace{5.5cm} + \Bigl[3 i T^7(x';x) \!+\! i T^{12}(x;x')\Bigr] \Phi(x') 
\Biggr\} . \qquad \label{onepotB}
\end{eqnarray}
The same comments apply to these results as for the mode equation (\ref{modeeqn}):
these equations are valid for any contribution to the graviton self-energy,
although what those contributions are in terms of the fundamental structure 
functions varies from matter to gravity according to relations (\ref{T5final}), 
(\ref{T7Final}) and (\ref{T10final}). And one should also note that derivatives
will be extracted from the structure functions, with primed ones partially 
integrated onto the potentials and unprimed ones taken outside the integration.

\section{Explicit Examples on de Sitter}

The previous section described our formalism for representing the graviton 
self-energy in cosmology. The purpose of this section is to put this formalism
in context with two explicit one loop results obtained on de Sitter background. 
As an example of matter contributions (\ref{mattersigma}) we consider the
dimensionally regulated result from a loop of massless, minimally coupled 
scalars \cite{Park:2011ww}. The more complex relation for gravity itself
(\ref{gravitonsigma}) is exemplified by an old $D=4$ computation 
\cite{Tsamis:1996qk} that was made in the simplest gauge \cite{Tsamis:1992xa}
before it was understood how to apply dimensional regularization.

\subsection{Contributions from a MMC Scalar}

Suppose that $S[\varphi,g]$ represents the sum of the scalar and gravitational 
actions, and that $\Delta S[g]$ stands for the counter-action. The one scalar 
loop contributions to the graviton self-energy can be expressed as the 
expectation value of the sum of three variational derivatives of these 
quantities,
\begin{eqnarray}
\lefteqn{ -i \Bigl[\mbox{}^{\mu\nu} \Sigma^{\rho\sigma}\Bigr](x;x') = 
\Biggl\langle \Omega \Biggl\vert T^*\Biggl[ \Bigl[\frac{i \delta S[\varphi,g]}{
\delta h_{\mu\nu}(x)} \Bigr]_{\varphi \varphi} \Bigl[ \frac{i \delta S[\varphi,g]}{
\delta h_{\rho\sigma}(x')} \Bigr]_{\varphi \varphi} } \nonumber \\
& & \hspace{3cm} + \Bigl[ \frac{i \delta^2 S[\varphi,g]}{\delta h_{\mu\nu}(x) 
\delta h_{\rho\sigma}(x')} \Bigr]_{\varphi \varphi} + \Bigl[ \frac{i \delta^2 
\Delta S[g]}{\delta h_{\mu\nu}(x) \delta h_{\rho\sigma}(x')} \Bigr]_{1} \Biggr] 
\Biggr\vert \Omega \Biggr\rangle , \qquad \label{operatorscalar}
\end{eqnarray}
where the subscripts indicate how many of the weak fields are retained and the
$T^*$-ordering symbol means any derivatives are taken after time ordering the
operators. Figure~\ref{diagrams} shows the associated Feynman diagrams.
\vskip .5cm
\begin{figure}[H]
\centering
\includegraphics[width=11cm]{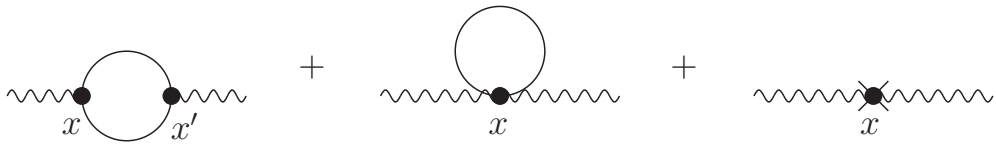}
\caption{\footnotesize One scalar loop contributions to the graviton self-energy, 
shown in the same order, left to right, as the three contributions 
to (\ref{operatorscalar}). Scalar lines are straight and graviton lines are wavy.}
\label{scalar}
\end{figure}

The 4-point diagram (the central one of Figure~\ref{scalar}) can be exactly 
canceled by the counterterm diagram (the right hand one of Figure~\ref{scalar}). 
The 3-point diagram (the left hand one of Figure~\ref{scalar}) takes the form
\cite{Park:2011ww,Park:2015kua},
\begin{eqnarray}
\lefteqn{-i \Bigl[\mbox{}^{\mu\nu} \Sigma^{\rho\sigma}_{\rm 3pt}\Bigr](x;x') = 
(a a')^{D+2} \Biggl\{ D^{\mu} {D'}^{(\rho} y \, {D'}^{\sigma)} D^{\nu} y \, 
\alpha(y) } \nonumber \\
& & \hspace{0.5cm} + D^{(\mu} y \, D^{\nu)} {D'}^{(\rho} y \, {D'}^{\sigma)} y \, 
\beta(y) + D^{\mu} y \, D^{\nu} y \, {D'}^{\rho} y \, {D'}^{\sigma} y \, 
\gamma(y) \nonumber \\
& & \hspace{1.5cm} + H^4 g^{\mu\nu} {g'}^{\rho\sigma} \, \delta(y) + H^2 \Bigl[ 
g^{\mu\nu} {D'}^{\rho} y \, {D'}^{\sigma} y + D^{\mu} y \, D^{\nu} y \, 
{g'}^{\rho\sigma} \Bigr] \epsilon(y) \Biggr\} . \qquad \label{scalar3pt} 
\end{eqnarray}
Here $y \equiv a a' H^2 \Delta x^2$ and its covariant derivatives are,
\begin{eqnarray}
D^{\mu} y = -\frac{H}{a} \Bigl[y \delta^{\mu}_{~0} \!-\! 2 a' H \Delta x^{\mu} 
\Bigr] \qquad , \qquad {D'}^{\rho} y = -\frac{H}{a'} \Bigl[y \delta^{\rho}_{~0}
\!+\! 2 a H \Delta x^{\rho}\Bigr] \; , \label{D1} \\
D^{\mu} {D'}^{\rho} y = \frac{H^2}{a a'} \Bigl[ y \delta^{\mu}_{~0} 
\delta^{\rho}_{~0} \!+\! 2 a \delta^{\mu}_{~0} H \Delta x^{\rho} \!-\! 2 H 
\Delta x^{\mu} a' \delta^{\rho}_{~0} \!-\! 2 \eta^{\mu\rho}\Bigr] \; .
\label{D2}
\end{eqnarray}
The various coefficients in expression (\ref{scalar3pt}) are given in terms of 
a single function $A(y)$ whose first derivative is \cite{Onemli:2002hr,
Onemli:2004mb},
\begin{eqnarray}
\lefteqn{A'(y) = -\frac{H^{D-2}}{4 (4 \pi)^{\frac{D}2}} \Biggl\{ \Gamma\Bigl(
\frac{D}2\Bigr) \Bigl( \frac{4}{y}\Bigr)^{\frac{D}2} + \Gamma\Bigl( \frac{D}2 
\!+\! 1\Bigr) \Bigl( \frac{4}{y}\Bigr)^{\frac{D}2 - 1} } \nonumber \\
& & \hspace{2cm} + \sum_{n=0}^{\infty} \Biggl[ \frac{\Gamma(n \!+\! 
\frac{D}2 \!+\! 2)}{\Gamma(n \!+\! 3)} \Bigl( \frac{y}{4}\Bigr)^{n-\frac{D}2 + 2} 
- \frac{\Gamma(n \!+\! D)}{\Gamma(n \!+\! \frac{D}2 \!+\! 1)} \Bigl( 
\frac{y}{4}\Bigr)^{n} \Biggr] \Biggr\} . \qquad \label{Aprime}
\end{eqnarray}
The functions $\alpha(y)$, $\beta(y)$, $\gamma(y)$, $\delta(y)$ and $\epsilon(y)$
are,\footnote{Note that the scalar functions $\alpha(y)$, 
$\beta(y)$ and $\gamma(y)$ are unrelated to the bi-scalar densities $\alpha(x;x')$,
$\beta(x;x')$ and $\gamma(x;x')$ defined in equations (\ref{gammadef}-\ref{betadef}).}
\begin{equation}
\alpha(y) = -\frac{\kappa^2}{2} {A'}^2 \;\; , \;\; \beta(y) = -\kappa^2 A' A'' 
\;\; , \;\; \gamma(y) = -\frac{\kappa^2}{2} {A''}^2 \; ,  
\label{scalaralpha}
\end{equation}
\begin{equation}
\delta(y) = -\frac{\kappa^2}{8} \Bigl[ (4 y \!-\! y^2)^2 {A''}^2 \!+\! 2 (2 \!-\! y)
(4 y \!-\! y^2) A' A'' \!+\! [4 (D\!-\! 4) \!-\! (4y \!-\! y^2)] {A'}^2 \Bigr] , 
\label{scalardelta}
\end{equation}
\begin{equation}
\epsilon(y) = \frac{\kappa^2}{4} \Bigl[ (4y \!-\! y^2) {A''}^2 + 2 (2 \!-\! y) 
A' A'' - {A'}^2 \Bigr] \; . \label{scalarepsilon}
\end{equation}

We can express (\ref{scalar3pt}) in the basis of Table~\ref{Tbasis} by first
$3+1$ decomposing the tensors of expressions (\ref{D1}-\ref{D2}),
\begin{equation}
\eta^{\mu\nu} = \overline{\eta}^{\mu\nu} - \delta^{\mu}_{~0} 
\delta^{\nu}_{~0} \qquad , \qquad \Delta x^{\mu} = \overline{\Delta x}^{\mu} 
+ \delta^{\mu}_{~0} \Delta \eta \; , \label{3+1}
\end{equation}
where $\Delta \eta \equiv x^0 - {x'}^0$.
Factors of $\overline{\Delta x}^{\mu}$ are then expressed as derivatives
using the rules,
\begin{eqnarray}
\overline{\Delta x}^{\alpha} f(\Delta x^2) = \frac{\overline{\partial}^{\mu}}{2}
I[f] & , & \overline{\Delta x}^{\alpha} \overline{\Delta x}^{\beta} f(\Delta x^2) 
= \frac{\overline{\partial}^{\alpha} \overline{\partial}^{\beta}}{4} I^2[f] \!-\! 
\frac{\overline{\eta}^{\alpha\beta}}{2} I[f] \; , \qquad \label{Rule1} \\
\overline{\Delta x}^{\alpha} \overline{\Delta x}^{\beta} \overline{\Delta x}^{\gamma}
f(\Delta x^2) & = & \frac18 \overline{\partial}^{\alpha} \overline{\partial}^{\beta}
\overline{\partial}^{\gamma} I^3[f] - \frac34 \overline{\eta}^{(\alpha \beta}
\overline{\partial}^{\gamma)} I^2[f] \; , \qquad \label{Rule2} \\
\overline{\Delta x}^{\alpha} \overline{\Delta x}^{\beta} \overline{\Delta x}^{\gamma}
\overline{\Delta x}^{\delta} f(\Delta x^2) & = & \frac1{16} \overline{\partial}^{\alpha} 
\overline{\partial}^{\beta} \overline{\partial}^{\gamma} \overline{\partial}^{\delta}
I^4[f] - \frac34 \overline{\eta}^{(\alpha \beta} \overline{\partial}^{\gamma}
\overline{\partial}^{\delta)} I^3[f] \nonumber \\
& & \hspace{4cm} + \frac34 \overline{\eta}^{(\alpha\beta} \overline{\eta}^{\gamma
\delta)} I^2[f] \; . \qquad \label{Rule3}
\end{eqnarray}
Here the operator $I[f]$ stands for the indefinite integral of $f(\Delta x^2)$ with 
respect to $\Delta x^2$. Table~\ref{Tscalar} gives the coefficient functions.

\begin{table}
\setlength{\tabcolsep}{8pt}
\def\arraystretch{1.5}
\centering
\begin{tabular}{|@{\hskip 1mm }c@{\hskip 1mm }||c|}
\hline
$i$ & $T^i(x;x')$ \\
\hline\hline
$1$ & $\delta - 2 (a^2 + {a'}^2) H^2 I[\epsilon] + 4 a^2 {a'}^2 
H^4 I^2[\gamma]$ \\
\hline
$2$ & $4 \alpha - 4 a a' H^2 I[\beta] + 8 a^2 {a'}^2 I^2[\gamma]$ \\
\hline
$3$ & $-\delta \!+\! (y \!+\! 2 a H \Delta \eta)^2 \epsilon$ \\
$$ & $- 2 {a'}^2 H^2 I[\alpha \!+\! (2 a H \Delta \eta \!+\! y) \beta \!+\! 
(4 a^2 H^2 \Delta \eta^2 \!+\! 4 a H \Delta \eta y \!+\! y^2) \gamma \!-\! 
\epsilon]$ \\
\hline
$5$ & $-2 a {a'}^2 H^3 I^2[\beta \!+\! 2 (2 a H \Delta \eta \!+\! y) \gamma]
+ 2 a H I[(2 a H \Delta \eta \!+\! y) \epsilon]$ \\
\hline
$7$ & $-2 a^2 {a'}^2 H^4 I^3[\gamma] + a^2 H^2 I^2[\epsilon]$ \\
\hline
$9$ & $-4 [2 \!+\! y \!+\! 2 a a' H^2 \Delta \eta^2] \alpha 
- 2 [y^2 \!-\! 2 a a' H^2 \Delta \eta^2 (2 \!-\! y)] \beta$ \\
$$ & $+ 2 a a' H^2 I[2 \alpha \!+\! (2 \!+\! 3 y) \beta \!+\! 4 a a' H^2 
\Delta \eta^2 \beta \!+\! 4 y^2 \gamma \!-\! 8 a a' H^2 \Delta \eta^2 
(2 \!-\! y) \gamma]$ \\
\hline
$10$ & $2 a H I[-2 \alpha \!+\! (2 a' H \Delta \eta \!-\! y) \beta] 
+ 4 a^2 a' H^3 I^2[\beta \!-\! 2 (2 a' H \Delta \eta \!-\! y) \gamma]$ \\
\hline
$12$ & $2 a a' H^2 I^2[\beta] - 8 a^2 {a'}^2 H^4 I^3[\gamma]$ \\
\hline
$$ & $(2 \!+\! y \!+\! 2 a a' H^2 \Delta \eta^2)^2 \alpha + 
(2 \!+\! y \!+\! 2 a a' H^2 \Delta \eta^2)$ \\
$13$ & $\times [y^2 \!-\! 2 a a' H^2 \Delta \eta^2 (2 \!-\! y)] \beta
+ (y^2 \!-\! 2 a a' H^2 \Delta \eta^2 (2 \!-\! y)]^2 \gamma$ \\
$$ & $+ \delta - 2 [y^2 \!+\! 2 a a' H^2 \Delta \eta^2 (2 \!+\! y)
+ 2 a^2 {a'}^2 H^4 \Delta \eta^4] \epsilon$ \\
\hline
$$ & $2 a H I[(2 \!+\! 2 a a' H^2 \Delta \eta^2) \alpha \!+\! y \alpha] 
\!-\! 4 a a' H^2 \Delta \eta (1 \!+\! a H \Delta \eta \!+\! a a' H^2 \Delta 
\eta^2) I[\beta]$ \\
$14$ & $+ 2 a H I[(1 \!-\! a' H \Delta \eta \!+\! 2 a a' H^2 \Delta \eta^2)
y \beta \!+\! y^2 \beta] + 8 a^2 {a'}^2 H^4 \Delta \eta^3 I[(2 \!-\! y) 
\gamma]$ \\
$$ & $-2 a H I[2 a a' H^2 \Delta \eta^2 (2 \!-\! y) y \gamma \!+\! (2 a' H
\Delta \eta \!-\! y) y^2 \gamma \!+\! (2 a H \Delta \eta \!+\! y) 
\epsilon]$ \\
\hline
$16$ & $H^2 I^2[a^2 \{\alpha \!-\! (2\!-\!y) \beta \!+\! (2\!-\!y)^2 \gamma
\!-\! \epsilon\} \!+\! a a' \{2\beta \!-\! 4 (2\!-\!y) \gamma\} \!+\! 4 {a'}^2 
\gamma]$ \\
\hline
$18$ & $a a' H^2 I^2[-2\alpha + 3 (2 - y) \beta - 32 \gamma + 4 (4 y - y^2) 
\gamma]$ \\
$$ & $+ (a^2 + {a'}^2) H^2 I^2[-4\beta + 8 (2 - y) \gamma]$ \\
\hline
$19$ & $a^2 a' H^3 I^3[-\beta + 2 (2 - y) \gamma] - 4 a {a'}^2 I^3[\gamma]$ \\
\hline
$21$ & $a^2 {a'}^2 H^4 I^4[\gamma]$ \\
\hline
\end{tabular}
\caption{\footnotesize Scalar contributions to the coefficient functions 
$T^{i}(x;x')$. The de Sitter length function is $y = a a' H^2 \Delta x^2$,
the various functions of it such as $\alpha(y)$ are defined in expressions 
(\ref{scalaralpha}-\ref{scalarepsilon}), and the operator ``$I$'' indicates 
indefinite integration with respect to $\Delta x^2$.}
\label{Tscalar}
\end{table}

It remains to comment on ultraviolet divergences and renormalization. 
Table~\ref{Tscalar} gives dimensionally regulated, primitive results. 
Comparing Table~\ref{Tscalar} with expressions (\ref{Aprime}) and 
(\ref{scalaralpha}-\ref{scalarepsilon}) reveals that the fundamental
structure functions have the following leading behaviors near coincidence:
\begin{equation}
T^{12} \sim \frac1{\Delta x^{2D-2}} \;\; , \;\; T^{16} \sim 
\frac1{\Delta x^{2D-2}} \;\; , \;\; T^{18} \sim \frac1{\Delta x^{2D-2}} \;\; 
, \;\; T^{19} \sim \frac{\Delta \eta}{\Delta x^{2D-2}} \; . \label{divergences}
\end{equation}
It must be recalled that the ultimate goal is to integrate $-i [\mbox{}^{\mu\nu}
\Sigma^{\rho\sigma}](x;x')$ of ${x'}$ in the quantum-corrected, linearized
Einstein equation (\ref{Einsteineqn}). Hence an expression such as 
$1/\Delta x^{2D-2}$ is quadratically divergent, while $1/\Delta x^{2D-4}$ is
logarithmically divergent. We localize these divergences by extracting 
derivatives until the integrable power of $1/\Delta x^{2D-6}$ is reached,
then adding zero in the form of the massless scalar propagator equation 
\cite{Onemli:2002hr},
\begin{eqnarray}
\lefteqn{\frac1{\Delta x^{2D-2}} = \frac{\partial^2}{2 (D \!-\!2)^2} 
\frac1{\Delta x^{2D-4}} = \frac{\partial^4}{4 (D\!-\!2)^2 (D\!-\!3) (D\!-\!4)}
\frac1{\Delta x^{2D-6}} \; , } \\
& & \hspace{0cm} = \frac{\partial^4}{4 (D\!-\!2)^2 (D\!-\!3) (D\!-\!4)} 
\Biggr[ \frac1{\Delta x^{2D-6}} \!-\! \frac{\mu^{D-4}}{\Delta x^{D-2}}\Biggr]
\nonumber \\
& & \hspace{6cm} + \frac{\mu^{D-4} \pi^{\frac{D}2} \partial^2 i 
\delta^D(x \!-\! x')}{(D\!-\!2)^2 (D\!-\!3) (D\!-\!4) \Gamma(\frac{D}2 \!-\! 1)} 
, \qquad \\
& & \hspace{0cm} = \frac{\mu^{D-4} \pi^{\frac{D}2} \partial^2 i 
\delta^D(x \!-\! x')}{(D\!-\!2)^2 (D\!-\!3) (D\!-\!4) \Gamma(\frac{D}2 \!-\! 1)}
- \frac{\partial^4}{32} \Biggl[\frac{\ln(\mu^2 \Delta x^2)}{\Delta x^2} \Biggr] 
+ O(D \!-\! 4) \; . \qquad \label{UV1}
\end{eqnarray}
We can similarly write,
\begin{equation}
\frac{\Delta \eta}{\Delta x^{2D-2}} = \frac{\mu^{D-4} \pi^{\frac{D}2} \partial_0 i 
\delta^D(x \!-\! x')}{(D\!-\!2) (D\!-\!3) (D\!-\!4) \Gamma(\frac{D}2 \!-\! 1)}
- \frac{\partial_0 \partial^2}{16} \Biggl[\frac{\ln(\mu^2 \Delta x^2)}{\Delta x^2} 
\Biggr] + O(D \!-\! 4) \; . \label{UV2}
\end{equation}

Renormalization is accomplished by using local counterterms to cancel the 
divergent delta functions in expressions (\ref{UV1}-\ref{UV2}). (This sometimes
leaves local residuals proportional to $\ln(a)$.) Because the derivatives on the 
contributions act on functions of $x^{\mu} - {x'}^{\mu}$, they can either be 
maintained as unprimed derivatives and pulled outside the ${x'}^{\mu}$ 
integration of the quantum-corrected Einstein equation (\ref{Einsteineqn}), or 
they can be reflected into primed derivatives ($\partial_{\mu} \rightarrow - 
\partial_{\mu}'$) and then partially integrated onto the graviton field 
$h_{\rho\sigma}(x')$.

\subsection{Contributions from Gravitons}

\begin{figure}[H]
\centering
\includegraphics[width=11cm]{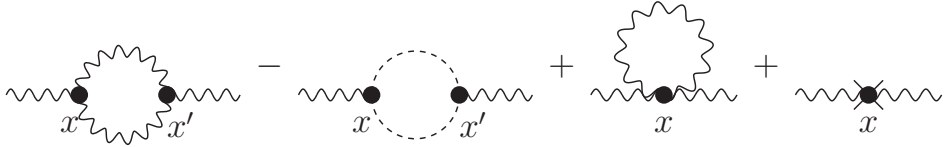}
\caption{\footnotesize Diagrams contributing to the one loop graviton 
self-energy, shown in the same order, left to right, as the three contributions 
to (\ref{operatorexpr}). Graviton lines are wavy and ghost lines are dashed.}
\label{diagrams}
\end{figure}

Suppose $S[g]$ represents the classical action of gravity, $S_g[h,\overline{\theta},
\theta]$ is the ghost and gauge fixing action, and $\Delta S[g]$ stands for the 
counter-action. The one loop graviton self-energy can be expressed as the 
expectation value of the sum of three variational derivatives of these quantities,
\begin{eqnarray}
\lefteqn{ -i \Bigl[\mbox{}^{\mu\nu} \Sigma^{\rho\sigma}\Bigr](x;x') = 
\Biggl\langle \Omega \Biggl\vert T^*\Biggl[ \Bigl[\frac{i \delta S[g]}{\delta 
h_{\mu\nu}(x)} \Bigr]_{h h} \Bigl[ \frac{i \delta S[g]}{\delta h_{\rho\sigma}(x')}
\Bigr]_{h h} + \Bigl[\frac{i \delta S[g]}{\delta 
h_{\mu\nu}(x)} \Bigr]_{\overline{\theta} \theta} } \nonumber \\
& & \hspace{0.3cm} \times \Bigl[ \frac{i \delta S[g]}{\delta h_{\rho\sigma}(x')}
\Bigr]_{\overline{\theta} \theta} + \Bigl[ \frac{i \delta^2 S[g]}{\delta 
h_{\mu\nu}(x) \delta h_{\rho\sigma}(x')} \Bigr]_{hh} + \Bigl[ \frac{i \delta^2 
\Delta S[g]}{\delta h_{\mu\nu}(x) \delta h_{\rho\sigma}(x')} \Bigr]_{1} \Biggr] 
\Biggr\vert \Omega \Biggr\rangle , \qquad \label{operatorexpr}
\end{eqnarray}
where the subscripts indicate how many graviton fields are retained and the
$T^*$-ordering symbol means any derivatives are taken after time ordering the
operators. Figure~\ref{diagrams} shows the associated Feynman diagrams.
\vskip .5cm

The actual computation \cite{Tsamis:1996qk} was made in $D=4$ dimensions before
it was understood how to employ dimensional regularization, so it can only be 
used away from coincidence. At the end of this section we discuss how it might
be extended to recover the full result. The computation was considerably more
difficult than deriving the scalar contribution of section 4.1. It differs from
the scalar result in three ways:
\begin{itemize}
\item{It breaks de Sitter invariance, both inessentially through the use of
a de Sitter breaking gauge \cite{Tsamis:1992xa,Woodard:2004ut} and unavoidably
through physical de Sitter breaking for gravitons \cite{Miao:2011fc,Faizal:2011iv,
Higuchi:2011vw,Miao:2011ng,Miao:2013isa} whose kinematics are the same as those 
of the massless, minimally coupled scalar \cite{Lifshitz:1945du,Allen:1987tz};}
\item{The coefficient functions $T^i(x;x')$ involve not only the two scale factors
and powers of $\Delta x^2$, but also up to a single factor of $\ln(H^2\Delta x^2)$; 
and}
\item{The self-energy is not annihilated by the action of a single Ward operator
so the coefficients $S^i(x;x')$ are nonzero.}
\end{itemize}   
The second point means that we can decompose the coefficient functions $T^i$
and $S^i$ into parts with and without a factor of $\ln(H^2 \Delta x^2)$,
\begin{eqnarray}
T^i(x;x') & = & T^i_{N}(x;x') + T^i_{L}(x;x') \!\times\! \ln(H^2 \Delta x^2) 
\; , \\
S^i(x;x') & = & S^i_{N}(x;x') + S^i_{L}(x;x') \!\times\! \ln(H^2 \Delta x^2) 
\; .
\end{eqnarray}
Tables~\ref{TN} and \ref{TL} give our results for the algebraically independent 
$T^i_N(x;x')$ and $T^i_L(x;x')$, respectively. Tables~\ref{SN} and \ref{SL} do
the $S^i_N(x;x')$ and $S^i_L(x;x')$.

\begin{table}
\setlength{\tabcolsep}{8pt}
\def\arraystretch{1.5}
\centering
\begin{tabular}{|@{\hskip 1mm }c@{\hskip 1mm }||c|}
\hline
$i$ & $T^i_{N}(x;x')$ \\
\hline\hline
$1$ & $\frac{8 (\frac{92}{5} - 77 a a' H^2 \Delta \eta^2 - 12 a^2 {a'}^2 
H^4 \Delta \eta^4)}{\Delta x^8} - \frac{\frac43 a a' (55 + 203 a a' H^2
\Delta \eta^2 + 48 a^2 {a'}^2 H^4 \Delta \eta^4) H^2}{\Delta x^6}$ \\
$$ & $- \frac{a^2 {a'}^2 (19 + 22 a a' H^2 \Delta \eta^2) H^4}{\Delta x^4}$ \\
\hline
$2$ & $\frac{\frac{1952}{5} - 416 a a' H^2 \Delta \eta^2}{\Delta x^8}
- \frac{\frac{16}{3} a a' (8 + 7 a a' H^2 \Delta \eta^2 - 6 a^2 {a'}^2
H^4 \Delta \eta^4) H^2}{\Delta x^6}$ \\
$$ & $+ \frac{4 a^2 {a'}^2 (14 + 3 a a' H^2 \Delta \eta^2) H^4}{\Delta x^4}$ \\
\hline
$3$ & $-\frac{\frac{64}{5} [23 + (55 a + 90 a') H \Delta \eta] \Delta \eta^2}{
\Delta x^{10}} - \frac{8 [23 + 7(16 a + 13 a') H \Delta \eta - (a^2 - 36 {a'}^2)
H^2 \Delta \eta^2]}{\Delta x^8}$ \\
$$ & $- \frac{4 [52 a^2 + 117 a a' - \frac{163}{3} {a'}^2 - 16 (a^3 - {a'}^3)
H \Delta \eta] H^2}{\Delta x^6} - \frac{a a' (18 a^2 + 147 a a' - 20 {a'}^2) H^4}{
\Delta x^4}$ \\
\hline
$5$ & $\frac{\frac{16}{5} [23 + (55 a + 90 a') H \Delta \eta] \Delta \eta}{\Delta x^8}
+ \frac{\frac{8}{3} [86 a + a' + (5 a^2 + 12 {a'}^2) H \Delta \eta] H}{\Delta x^6}
- \frac{2 a (6 a^2 - 33 a a' + 4 {a'}^2) H^3}{\Delta x^4}$ \\
\hline
$7$ & $-\frac{\frac{4}{15} [23 + (55 a + 90 a') H \Delta \eta]}{\Delta x^6}
+ \frac{( \frac13 a^2 - 6 a a' + 8 {a'}^2) H^2}{\Delta x^4} + \frac{ a a'
(2 a^2 - 6 a a' + {a'}^2) H^4}{\Delta x^2}$ \\
\hline
$9$ & $-\frac{\frac{128}{5} (61 - 55 a a' H^2 \Delta \eta^2) \Delta \eta^2}{\Delta x^{10}}
- \frac{16 (61 + 10 a a' H^2 \Delta \eta^2 - 2 a^2 {a'}^2 H^4 \Delta \eta^4)}{\Delta x^8}$ \\
$$ & $- \frac{8 a a' (70 - a a' H^2 \Delta \eta^2) H^2}{\Delta x^6}
-\frac{12 a^2 {a'}^2 H^4}{\Delta x^4}$ \\
\hline
$10$ & $\frac{\frac{16}{5} (61 - 40 a a' H^2 \Delta \eta^2) \Delta \eta}{\Delta x^8}
+ \frac{\frac{16}{3} [20 a - 16 a' + (9a - 2a') a' H \Delta \eta] H}{\Delta x^6}
+ \frac{2 a' (a^2 - 19 a a' - 2 {a'}^2) H^3}{\Delta x^4}$ \\
\hline
$12$ & $-\frac{\frac{8}{15} (61 - 20 a a' H^2 \Delta \eta^2)}{\Delta x^6}
- \frac{\frac23 a a' ( 16 + 5 a a' H^2 \Delta \eta^2) H^2}{\Delta x^4}
+ \frac{2 a^2 {a'}^2 (4 - a a' H^2 \Delta \eta^2) H^4}{\Delta x^2}$ \\
\hline
$13$ & $\frac{5376 \Delta \eta^4}{\Delta x^{12}} + \frac{64 (84 + 21 a a' H^2 
\Delta \eta^2) \Delta \eta^2}{\Delta x^{10}} + \frac{8 (126 + 434 a a' H^2 
\Delta \eta^2 + 53 a^2 {a'}^2 H^4 \Delta \eta^4)}{\Delta x^8}$ \\
$$ & $+ \frac{4 a a' (409 + 36 a a ' H^2 \Delta \eta^2 - 24 a^2 {a'}^2 H^4
\Delta \eta^4) H^2}{\Delta x^6} + \frac{a^2 {a'}^2 (557 + 24 a a' H^2
\Delta \eta^2) H^4}{\Delta x^4}$ \\
\hline
$14$ & $-\frac{\frac{5376}{5} \Delta \eta^3}{\Delta x^{10}}
- \frac{16 [42 + (19 a - 15 a') H \Delta \eta] \Delta \eta}{\Delta x^8}$ \\
$$ & $+ \frac{8 [-\frac{238}{3} a + 56 a' + \frac{1}{3} (13a^2+48a'^2) H \Delta \eta
- 6a a' (3 a \!-\! 4 a') H^2 \Delta \eta^2] H}{\Delta x^6}
+ \frac{2 a (8 a^2 - 139 a a' -6 {a'}^2) H^3}{\Delta x^4}$ \\
\hline
$16$ & $\frac{\frac{336}{5} \Delta \eta^2}{\Delta x^8}
+ \frac{4 [\frac{13}{3} - (a + 6 a') H \Delta \eta)]}{\Delta x^6}
+ \frac{(\frac{67}{3} a^2 + 14 a a' + 6 {a'}^2) H^2}{\Delta x^4}
- \frac{a a' (2 a^2 - 10 a a' + 3 {a'}^2) H^4}{\Delta x^2}$ \\
\hline
$18$ & $\frac{\frac{1344}{5} \Delta \eta^2}{\Delta x^8}
+ \frac{8 (\frac{29}{3} + 6 a a' H^2 \Delta \eta^2)}{\Delta x^6}
+ \frac{2 a a' (-46 + 5 a a' H^2 \Delta \eta^2) H^2}{\Delta x^4}$ \\
\hline
$19$ & $-\frac{\frac{112}{5} \Delta \eta}{\Delta x^6} + \frac{(\frac{26}{3} a + 8 a') 
H}{\Delta x^4} - \frac{a^2 {a'}^2 H^4 \Delta \eta}{\Delta x^2}$ \\
\hline
$21$ & $\frac{\frac{14}{5}}{\Delta x^4}$ \\
\hline
\end{tabular}
\caption{\footnotesize Contributions to $T^{i}(x;x')$ that do not contain factors
of $\ln(H^2 \Delta x^2)$. Each of the tabulated terms must be multiplied by 
$-\frac{\kappa^2}{64 \pi^4}$.}
\label{TN}
\end{table}

\begin{table}
\setlength{\tabcolsep}{8pt}
\def\arraystretch{1.5}
\centering
\begin{tabular}{|@{\hskip 1mm }c@{\hskip 1mm }||c|}
\hline
$i$ & $T^i_{L}(x;x')$ \\
\hline\hline
$1$ & 
$\frac{16 a a' (a^2 + {a'}^2) H^4 \Delta \eta^2}{\Delta x^6} - \frac{4 a a' (a^2 - 
4 a a' + {a'}^2) H^4}{\Delta x^4} + \frac{12 a^3 {a'}^3 H^6}{\Delta x^2}$ \\
\hline
$2$ & 
$-\frac{32 a a' (a^2 + {a'}^2) H^4 \Delta \eta^2}{\Delta x^6} - \frac{16 a^2 {a'}^2
H^4}{\Delta x^4} - \frac{4 a^3 {a'}^3 H^6}{\Delta x^2}$ \\
\hline
$3$ & 
$-\frac{16 a^3 {a'}^2 H^5 \Delta \eta^3}{\Delta x^6} + \frac{4 a^2 {a'}^2
(3 a - 2 a') H^5 \Delta \eta}{\Delta x^4} - \frac{16 a^3 {a'}^3 H^6}{
\Delta x^2}$ \\
\hline
$5$ & 
$\frac{8 a^2 a' (a + a') H^4 \Delta \eta}{\Delta x^4} + 
\frac{12 a^3 {a'}^2 H^5}{\Delta x^2}$ \\
\hline
$7$ & 
$-\frac{2 a a' (2 a^2 + a a' + {a'}^2) H^4}{\Delta x^2}$ \\
\hline
$9$ &
$-\frac{1536 a a' H^2 \Delta \eta^4}{\Delta x^{10}} - \frac{96 (a^2 +10 a a' + {a'}^2)
H^2 \Delta \eta^2}{\Delta x^8} - \frac{32 (a^2 + a a' + {a'}^2) H^2}{\Delta x^6}
+ \frac{4 a^2 {a'}^2 H^4}{\Delta x^4}$ \\
\hline
$10$ &  
$\frac{192 a a' H^2 \Delta \eta^3}{\Delta x^8} + \frac{16 (a^2 + 2 a a' + 3 {a'}^2)
H^2 \Delta \eta}{\Delta x^6} - \frac{4 a' (a^2 - a a' + 2 {a'}^2) H^3}{\Delta x^4} 
+ \frac{4 a^2 {a'}^3 H^5}{\Delta x^2}$ \\
\hline
$12$ & 
$-\frac{32 a a' H^2 \Delta \eta^2}{\Delta x^{6}} - \frac{4 (3 a^2 - 4 a a' + 3 {a'}^2) 
H^2}{\Delta x^4} + \frac{2 a a' (2 a^2 + 5 a a' + 2 {a'}^2) H^4}{\Delta x^2}$ \\
\hline
$13$ & 
$-\frac{1536 a a' H^2 \Delta \eta^4}{\Delta x^{10}} + \frac{96 (a^2 - 14 a a' + {a'}^2)
H^2 \Delta \eta^2}{\Delta x^8} + \frac{32 a a' (-3 + 7 a a' H^2 \Delta \eta^2 +
a^2 {a'}^2 H^4 \Delta \eta^4) H^2}{\Delta x^6}$ \\
$$ & $- \frac{4 a a' (4 a^2 - 19 a a' + 4 {a'}^2)
H^4}{\Delta x^4} + \frac{24 a^3 {a'}^3 H^6}{\Delta x^2}$ \\
\hline
$14$ & 
$\frac{384 a a' H^2 \Delta \eta^3}{\Delta x^8} 
- \frac{32 (a^2 - 6 a a' - {a'}^2) H^2 \Delta \eta}{\Delta x^6} 
-\frac{8 a (2a^2 + a a' - {a'}^2) H^3}{\Delta x^4} 
- \frac{16 a^3 {a'}^2 H^5}{\Delta x^2}$ \\
\hline
$16$ &
$-\frac{32 a a' H^2 \Delta \eta^2}{\Delta x^6} + \frac{4 (a^2 - 3 {a'}^2) H^2}{
\Delta x^4} + \frac{2 a a' (2 a^2 - {a'}^2) H^4}{\Delta x^2}$ \\
\hline
$18$ & 
$-\frac{96 a a' H^2 \Delta \eta^2}{\Delta x^{6}} - \frac{12 (a^2 + {a'}^2) H^2
}{\Delta x^4} - \frac{10 a^2 {a'}^2 H^4}{\Delta x^2}$ \\
\hline
$19$ & $\frac{8 a a' H^2 \Delta \eta}{\Delta x^{4}} + 
\frac{2 a a' (a - 3 a') H^3}{\Delta x^2}$ \\
\hline
$21$ & $0$ \\
\hline
\end{tabular}
\caption{\footnotesize Contributions to $T^{i}(x;x')$ that contain factors
of $\ln(H^2 \Delta x^2)$. Each tabulated term must be multiplied by $-
\frac{\kappa^2}{64 \pi^4}$.}
\label{TL}
\end{table}

\begin{table}
\setlength{\tabcolsep}{8pt}
\def\arraystretch{1.5}
\centering
\begin{tabular}{|@{\hskip 1mm }c@{\hskip 1mm }||c|}
\hline
$i$ & $S^i_{N}(x;x')$ \\
\hline\hline
$$ & $\frac{2560 a' H \Delta \eta^4}{\Delta x^{12}} 
+ \frac{128 [42 a - 6 a' + (9 a^2 + 8 {a'}^2) H \Delta \eta] H \Delta \eta^2}{
\Delta x^{10}}$ \\
$1$ & $+ \frac{16 [303 a - 193 a' + (73 a^2 - 33 {a'}^2) H \Delta \eta
+ (12 a^3 + 18 {a'}^3) H^2 \Delta \eta^2] H}{\Delta x^8}$ \\
$$ & $+ \frac{8 (20 a^3 + 41 a^2 a' + 50 a {a'}^2 - 14 {a'}^3) H^3}{\Delta x^6} 
+ \frac{8 a^2 {a'}^2 (5 a + 4 a') H^5}{\Delta x^4}$ \\
\hline
$2$ & $-\frac{256 a' H \Delta \eta^3}{\Delta x^{10}} - \frac{16 [(42 a - 16 a') 
H \Delta \eta + (9 a^2 + 6 {a'}^2) H^2 \Delta \eta^2]}{\Delta x^8}$ \\
$$ & $- \frac{ 8 H^2 (17 a^2 + \frac{53}{3} a a' - 5 {a'}^2 + 4 a^3 H \Delta \eta)
}{\Delta x^6} - \frac{2 a^2 a' H^4 (5 a' + 12 a)}{\Delta x^4}$ \\
\hline
$3$ & $\frac{-384 (a^2-3aa'-a'^2) H^2 \Delta \eta^3}{
\Delta x^{10}} - \frac{16 [57 a + 7 a' - (5 a^2 - {a'}^2) H \Delta \eta + 
12 {a'}^3 H^2 \Delta \eta^2] H}{\Delta x^8}$ \\
$$ & $+ \frac{8 (2 a^3 - \frac{65}{3} a {a'}^2 + 9 {a'}^3) H^3}{\Delta x^6}
- \frac{2 a^2 {a'}^2 (4 a + 11 a') H^5}{\Delta x^4}$ \\
\hline
$4$ & $\frac{32 a' H \Delta \eta^2}{\Delta x^8} + \frac{8 H [11 a - 5 a' + (4 a^2 + 
\frac13 {a'}^2) H \Delta \eta]}{\Delta x^6}$ \\
$$ & $+ \frac{2 H^3 a [4 a^2 + 2 a a' - \frac{11}{3} {a'}^2]}{\Delta x^4} + 
\frac{H^5 a^2 {a'}^2 (4 a \!+\! a')}{\Delta x^2}$ \\
\hline
$5$ & $-\frac{256 a' H \Delta \eta^3}{\Delta x^{10}} + \frac{16 [6a - 16 a'
+ (3 a^2 - 8 {a'}^2) H \Delta \eta] H \Delta \eta}{\Delta x^8}$ \\
$$ & $+ \frac{8 [7 a^2 - a a' - \frac{11}{3} {a'}^2 - 6 {a'}^3 H \Delta \eta] H^2}{
\Delta x^6} - \frac{4 a {a'}^2 (a - 2 a') H^4}{\Delta x^4}$ \\
\hline
$6$ & $-\frac{256 (3 a - 2 a') H \Delta \eta^2}{\Delta x^{10}}
- \frac{4 [90 a - 50 a' + (27a^2 + 3 {a'}^2) H\Delta \eta] H}{\Delta x^8}
- \frac{6 a a' (9a - a') H^3}{\Delta x^6}$ \\
\hline
$7$ & $\frac{32 a' H \Delta \eta^2}{\Delta x^8} - \frac{8 H (a^3 - a^2 a' - 
\frac{10}{3} a {a'}^2 + 2 {a'}^3)}{a a' \Delta x^6} + \frac{H^3 a a' (a - 5 a')
}{2 \Delta x^4}$ \\
\hline
$8$ & $\frac{32 (3 a - 2 a') H \Delta \eta}{\Delta x^8} 
+ \frac{(18 a^2 - 20 a a' + \frac{38}{3} {a'}^2) H^2}{\Delta x^6}
+ \frac{a {a'}^2 (a' - \frac92 a) H^4}{\Delta x^4}$ \\
\hline
$9$ & $\frac{48(a^2 - 3 a a' - {a'}^2) H^2 \Delta \eta^2}{\Delta x^8}
+ \frac{8 (-2a^2 + 2 a a' - 3 {a'}^2 + 4 {a'}^3 H \Delta \eta) H^2}{\Delta x^6}
+ \frac{2 a {a'}^2 (2 a - 3 a') H^4}{\Delta x^4}$ \\
\hline
$10$ & $-\frac{16 a' H \Delta \eta}{3 \Delta x^6} + \frac{2 H^2 a'(a - \frac13 a')
}{\Delta x^4} + \frac{H^4 a^2 {a'}^2}{2 \Delta x^2}$ \\
\hline
\end{tabular}
\caption{\footnotesize Contributions to $S^i(x;x')$ that do not contain a factor
of $\ln(H^2 \Delta x^2)$. Each of the tabulated terms must be multiplied by $-
\frac{\kappa^2}{64\pi^4}$.}
\label{SN}
\end{table}

\begin{table}
\setlength{\tabcolsep}{8pt}
\def\arraystretch{1.5}
\centering
\begin{tabular}{|@{\hskip 1mm }c@{\hskip 1mm }||c|}
\hline
$i$ & $S^i_{L}(x;x')$ \\
\hline\hline
$1$ & $\frac{384 a^2 {a'}^2 H^4 \Delta \eta^3}{\Delta x^8}
+ \frac{64 a a' (a^2 + 5 a a' - 3 {a'}^2) H^4 \Delta \eta}{\Delta x^6}
+ \frac{64 a^3 {a'}^2 H^5}{\Delta x^4}$ \\
\hline
$2$ & $-\frac{64 a^2 {a'}^2 H^4 \Delta \eta^2}{\Delta x^6}
- \frac{16 a^3 a' H^4}{\Delta x^4}$ \\
\hline
$3$ & $\frac{192 a^2 {a'}^2H^4 \Delta \eta^3}{\Delta x^8}
+ \frac{32 aa'(a^2 - 2 a a' + 4 {a'}^2) H^4 \Delta \eta}{\Delta x^6}
- \frac{24 a^3 {a'}^2 H^5}{\Delta x^4}$ \\
\hline
$4$ & $\frac{8 a^2 {a'}^2 H^4 \Delta \eta}{\Delta x^4} + 
\frac{4 a^3 {a'}^2 H^5}{\Delta x^2}$ \\
\hline
\end{tabular}
\caption{\footnotesize Nonzero parts of $S^i(x;x')$ which are proportional to 
$\ln(H^2 \Delta x^2)$. (The cases of $i = 5,6,7,8,9,10$ vanish.) Each of the 
tabulated terms must be multiplied by $-\frac{\kappa^2}{64 \pi^4}$.}
\label{SL}
\end{table}

We also need the auxiliary functions defined in expressions (\ref{gammadef}),
(\ref{alphadef}) and (\ref{betadef}). The antisymmetric part of $\gamma(x;x')$ 
is,
\begin{eqnarray}
\lefteqn{\gamma - \gamma^{R} = (\eta \!+\! \eta') \Delta \eta \Biggl\{
-\frac{5376 a a' H^2 \Delta \eta^2}{5 \Delta x^{10}} - \frac{8 a a' H^2 
[359 \!+\! 124 a a' H^2 \Delta \eta^2]}{5 \Delta x^8} } \nonumber \\
& & \hspace{-0.5cm} - \frac{4 a^2 {a'}^2 H^4 [181 \!+\! 60 a a' H^2 \Delta \eta^2]}{
15 \Delta x^6} + \frac{a^3 {a'}^3 H^6 [24 \!+\! 11 a a' H^2 \Delta \eta^2]}{
\Delta x^4} \nonumber \\
& & \hspace{-0.5cm} - \frac{2 a^4 {a'}^4 H^8 [5 \!-\! 2 a a' H^2 \Delta \eta^2]}{
\Delta x^2} + \Biggl[ \frac{96 a^2 {a'}^2 H^4 \Delta \eta^2}{\Delta x^8}
+ \frac{8 a^2 {a'}^2 H^4 [6 \!+\! 7 a a' H^2 \Delta \eta^2]}{\Delta x^6}
\nonumber \\
& & \hspace{-0.5cm} + \frac{2 a^3 {a'}^3 H^6 [1 \!+\! 7 a a' H^2 \Delta \eta^2]}{
\Delta x^4} - \frac{a^4 {a'}^4 H^8 [31 \!+\! 8 a a' H^2 \Delta \eta^2]}{
\Delta x^2} \Biggr] \ln(H^2 \Delta x^2) \Biggl\} . \qquad \label{gammaanti}
\end{eqnarray}
This implies that $\alpha(x;x')$ and $\beta(x;x')$ are,
\begin{eqnarray}
\lefteqn{\alpha = \frac{\frac{672}{5} \Delta \eta^2}{\Delta x^8} +
\frac{\frac{764}{15} \!+\! \frac{16}{3} a a' H^2 \Delta \eta^2}{\Delta x^6}
- \frac{a a' H^2 [\frac{16}3 \!+\! \frac53 a a' H^2 \Delta \eta^2]}{\Delta x^4} }
\nonumber \\
& & \hspace{0cm}  + \frac{a^2 {a'}^2 H^4 [4 \!-\! a a' H^2 \Delta \eta^2]}{
\Delta x^2} + \Biggl[- \frac{16 a a'H^2 \Delta \eta^2}{\Delta x^6} 
-\frac{2 a a' H^2 [2 \!+\! 3 a a' H^2 \Delta \eta^2]}{\Delta x^4} \nonumber \\
& & \hspace{4.5cm} + \frac{a^2 {a'}^2 H^4 [9 \!+\! 2 a a' H^2 \Delta \eta^2]}{
\Delta x^2} \Biggr] \ln(H^2 \Delta x^2) , \qquad \label{gravalpha} \\
\lefteqn{\beta = -\frac{[\frac{3904}{5} \!-\! 704 a a' H^2 \Delta \eta^2]
\Delta \eta^2}{\Delta x^{10}} - \frac{[\frac{488}{5} \!+\! 272 a a' H^2 
\Delta \eta^2 \!-\! 80 a^2 {a'}^2 H^4 \Delta \eta^4]}{\Delta x^8} }
\nonumber \\
& & \hspace{-0.5cm} - \frac{a a' H^2 [\frac{280}{3} \!+\! \frac{332}{3} a a'
H^2 \Delta \eta^2]}{\Delta x^6} \!+\! \frac{4 a^2 {a'}^2 H^4 (6 \!-\! a a' H^2
\Delta \eta^2)}{\Delta x^4} \!+\! \Biggl[ -\frac{768 a a' H^2 \Delta \eta^4}{
\Delta x^{10}} \nonumber \\
& & \hspace{-0.5cm} - \frac{144 a a' H^2 \Delta \eta^2 [4 \!+\! a a' H^2 \Delta 
\eta^2]}{\Delta x^8} - \frac{8 a a' H^2 [6 \!+\! 8 a a' H^2 \Delta \eta^2
\!+\! 2 a^2 {a'}^2 H^4 \Delta \eta^4]}{\Delta x^6} \nonumber \\
& & \hspace{2.9cm} + \frac{2 a^2 {a'}^2 [1 \!+\! 2 a a' H^2 \Delta \eta^2]}{
\Delta x^4} - \frac{4 a^3 {a'}^3 H^6}{\Delta x^2} \Biggr] \ln(H^2 \Delta x^2)
. \qquad \label{gravbeta}
\end{eqnarray}
Note that none of the structure functions differ in form from the primitive 
contributions. That is not some miracle of the de Sitter background; one can
see that it must be true generally from equations (\ref{T2final}) --- which
could be used to infer $\beta(x;x')$ --- and equation (\ref{T7Final}) --- 
which could be used to infer $\alpha(x;x')$. The absence of new functional
forms is quite unlike what happened in previous representations of the scalar 
result \cite{Park:2011ww,Park:2015kua}, neither of which could even be applied 
to contributions from gravity or to general cosmological backgrounds.

\subsubsection{Recovering the Local Terms}

The previous results determine the nine structure functions for all ${x'}^{\mu} 
\neq x^{\mu}$. However, there are still potentially important local contributions 
proportional to $i\delta^4(x - x')$. These terms dominate the fermion wave 
function \cite{Miao:2005am,Miao:2006gj} and the photon field strength 
\cite{Leonard:2013xsa,Wang:2014tza}, and they make an important contribution to 
electromagnetic forces \cite{Glavan:2013jca}, so it is worth explaining how 
they can be recovered. Of course we could simply re-do the computation using
dimensional regularization from the beginning using the $D$-dependent propagators
\cite{Woodard:2004ut} and vertices \cite{Tsamis:1992xa}, but we have in mind a
simpler approach based on understanding the three sources of local 
contributions:\footnote{The
reader is free to dismiss the comments of this subsection as conjectural. However,
they are based on the authors' great familiarity with the $D=4$ computation of
the graviton loop contribution \cite{Tsamis:1996qk}, and the close simularity
of that contribution to the general $D$ scalar loop contribution \cite{Park:2011ww}
reported in section 4.1. In particular, the graviton 3-point vertex takes the
same $\kappa a^{D-2} h \partial h \partial h$ form as the scalar-graviton vertex
$\kappa a^{D-2} h \partial \varphi \partial \varphi$. The most important part
of the $D$-dimensional graviton propagator \cite{Tsamis:1992xa,Woodard:2004ut} 
is also just some constant tensors times the scalar propagator.}
\begin{itemize}
\item{From renormalization;}
\item{From the 4-point diagram of Figure~\ref{diagrams}; and}
\item{From differentiated propagators in the two 3-point diagrams of 
Figure~\ref{diagrams}.}
\end{itemize}

Renormalization is the simplest case to understand. From Tables~\ref{TN} and 
\ref{TL} we see that the most singular parts of the fundamental coefficient 
functions near coincidence are,
\begin{equation}
T^{12} \sim \frac1{\Delta x^6} \;\; , \;\; T^{16} \sim \frac1{\Delta x^6} \;\;
, \;\; T^{18} \sim \frac1{\Delta x^6} \;\; , \;\; T^{19} \sim 
\frac{\Delta \eta}{\Delta x^6} \; . \label{gravitydiv}
\end{equation}
In general $D$ the ghost and graviton propagators involve functions of the
same form as (\ref{Aprime}) in the scalar propagator \cite{Woodard:2004ut},
\begin{equation}
i\Delta(x;x') \sim \Bigl( \frac1{a a' \Delta x^2}\Bigr)^{\frac{D}2 -1}
+ \Bigl( \frac1{a a' \Delta x^2}\Bigr)^{\frac{D}2 - 2} + \dots 
\end{equation}
The generic vertex involves a factor of $a^{D-2}$ with two derivatives
\cite{Tsamis:1992xa}. So comparison with the leading scalar divergences
(\ref{divergences}) means that $D=4$ results like $1/\Delta x^6$ correspond to
$1/\Delta x^{2D-2}$ in the dimensionally regulated theory, and all $D$-dependent
powers of the scale factor cancel between the two propagators and the two
vertices. Hence we can extend the $D=4$ results from Tables~\ref{TN}, \ref{TL}, 
\ref{SN} and \ref{SL} by the replacements such as,
\begin{eqnarray}
\frac1{\Delta x^6} & \longrightarrow & \frac{\mu^{D-4} \pi^{\frac{D}2} 
\partial^2 i \delta^D(x \!-\! x')}{(D\!-\!2)^2 (D\!-\!3) (D\!-\!4)
\Gamma(\frac{D}2 \!-\! 1)} - \frac{\partial^4}{32} \Biggl[ \frac{\ln(\mu^2
\Delta x^2)}{\Delta x^2}\Biggr] \; , \qquad \\
\frac{\Delta \eta}{\Delta x^6} & \longrightarrow & \frac{\mu^{D-4} \pi^{\frac{D}2} 
\partial_0 i \delta^D(x \!-\! x')}{(D\!-\!2) (D\!-\!3) (D\!-\!4)
\Gamma(\frac{D}2 \!-\! 1)} - \frac{\partial_0 \partial^2}{16} \Biggl[ 
\frac{\ln(\mu^2 \Delta x^2)}{\Delta x^2}\Biggr] \; , \qquad
\end{eqnarray}
times non-negative integer powers of $a$ and $a'$. Because counterterms are
proportional to $a^{D-4} i \delta^D(x - x')$ times non-negative powers of $a$ and
$a'$, it is possible to predict the finite factors of $\ln(a) \delta^4(x - x')$
that remain after renormalization. Note that we can also predict how to extract
derivatives from the finite, nonlocal parts of the structure functions.

All the 4-point diagrams are local, and they are simple enough to compute 
directly. In dimensional regularization any $D$-dependent power of $\Delta x^2$ 
vanishes at coincidence. The coincident propagator comes from the integer 
sums in (\ref{Aprime}) and related propagator functions. These will never 
contribute $D$-dependent powers of $a$, hence the factor of $a^{D-2}$ from the
4-point agrees with the $D$-dependent factor of $a$ from the counterterm, so
no finite factors of $\ln(a)$ can arise from this source. However, the 
coincident propagator can produce an easily-predictable and ultraviolet finite 
factor of $\ln(a)$.

Without regard to the tensor structure we can see that the generic 3-point 
contribution takes the form,
\begin{equation}
\Bigl(i\kappa a^{D-2} \partial^2\Bigr) \times i\Delta(x;x') i\Delta(x;x') 
\times \Bigr(i \kappa {a'}^{D-2} {\partial'}^2\Bigr) \; . 
\end{equation}
Acting two times derivatives on a propagator produces a delta function 
\cite{Onemli:2002hr},
\begin{equation}
\partial_{\mu} \partial_{\nu}' i\Delta(x;x') = \frac{\delta^0_{~\mu} 
\delta^0_{~\nu} i \delta^D(x \!-\! x')}{a^{D-2}} + {\rm Nonlocal\ Terms} \; .
\end{equation}
The other propagator is taken to coincidence by the delta function, so the 
same considerations apply to it as for the 4-point contributions considered
above. It turns out that acting the derivatives to produce the nonlocal
terms is the rate-limiting step of the computation, so it is considerably 
simpler to access the local $\ln(a)$ term than to derive the dimensionally
regulated nonlocal contributions.

\subsubsection{The Gauge Issue}

Graviton propagators require gauge fixing. The calculation \cite{Tsamis:1996qk}
reported in section 4.2 was performed by adding a gauge fixing functional whose 
$D$-dimensional extension is \cite{Tsamis:1992xa,Woodard:2004ut},
\begin{equation}
\mathcal{L}_{GF} = -\frac{a^{D-2}}{2} \eta^{\mu\nu} F_{\mu} F_{\nu} \;\; , \;\;
F_{\mu} = \eta^{\rho\sigma} \Bigl( h_{\mu\rho , \sigma} - \frac12 h_{\rho\sigma , \mu}
+ (D\!-\! 2) a H h_{\mu\rho} \delta^{0}_{~\sigma} \Bigr) \; . \label{dSgauge}
\end{equation}
The special feature of this gauge is that it makes the propagator take the 
form of a sum of three constant tensor factors times scalar propagators,
\begin{equation}
i\Bigl[\mbox{}_{\mu\nu} \Delta_{\rho\sigma}\Bigr](x;x') = \sum_{I=A,B,C} 
\Bigl[\mbox{}_{\mu\nu} T^I_{\rho\sigma}\Bigr] \times i\Delta_I(x;x') \; .
\end{equation}
Here the constant tensor factors are,
\begin{eqnarray} 
\Bigl[\mbox{}_{\mu\nu} T^A_{\rho\sigma}\Bigr] = 2 \overline{\eta}_{\mu (\rho}
\overline{\eta}_{\sigma) \nu} - \frac{2}{D\!-\!3} \overline{\eta}_{\mu\nu}
\overline{\eta}_{\rho\sigma} \quad , \quad \Bigl[\mbox{}_{\mu\nu} T^B_{\rho\sigma}
\Bigr] = -4 \delta^0_{~(\mu} \overline{\eta}_{\nu) (\rho} \delta^0_{~ \sigma)} 
\; , \qquad \\
\Bigl[\mbox{}_{\mu\nu} T^C_{\rho\sigma}\Bigr] = \frac{2 E_{\mu\nu} E_{\rho\sigma}}{
(D\!-\!2) (D\!-\!3)} \quad , \quad E_{\mu\nu} \equiv (D\!-\!3) \delta^0_{~\mu}
\delta^0_{~\nu} + \overline{\eta}_{\mu\nu} \; . \qquad 
\end{eqnarray}
And the three scalar propagators are all related to the function $A(y)$ of
expression (\ref{Aprime}),
\begin{eqnarray}
i\Delta_A(x;x') & = & A(y) + k \ln(a a') \qquad k \equiv 
\frac{H^{D-2}}{(4\pi)^{\frac{D}2}} \frac{\Gamma(D\!-\!1)}{\Gamma(\frac{D}2)} \; , \\
i\Delta_B(x;x') & = & B(y) \equiv -\frac{[(4 y \!-\! y^2) A'(y) \!+\! (2 \!-\! y) k]}{
2 (D \!-\! 2)} \; , \\
i\Delta_C(x;x') & = & C(y) \equiv \frac12 (2 \!-\! y) B(y) + \frac{k}{D\!-\!3} \; .
\end{eqnarray}

The flat space limit is obtained by taking the scale factor to unity and the 
Hubble parameter to zero. In this limit our de Sitter gauge reduces to,
\begin{equation}
\mathcal{L}_{GF} \longrightarrow -\frac12 \eta^{\mu\nu} F_{\mu} F_{\nu} \qquad , 
\qquad F_{\mu} = \eta^{\rho\sigma} \Bigl( h_{\mu\rho , \sigma} - \frac12 
h_{\rho\sigma , \mu} \Bigr) \; , \label{flatgauge}
\end{equation}
and the corresponding propagator becomes,
\begin{equation}
i\Bigl[\mbox{}_{\mu\nu} \Delta_{\rho\sigma}\Bigr](x;x') \longrightarrow 
\Bigl( 2 \eta_{\mu (\rho} \eta_{\sigma ) \nu} - \frac{2}{D\!-\! 2} \eta_{\mu\nu}
\eta_{\rho\sigma} \Bigr) \times i\Delta(x;x') \; ,
\end{equation}
where the massless scalar propagator of flat space was defined in expression 
(\ref{flatprop}). This is precisely the gauge Capper employed to derive the
results reported in equations (\ref{flatsigma1}-\ref{T4}) \cite{Capper:1979ej}.
It is straightforward to check that the flat space limits of the de Sitter 
results reported in Tables~\ref{TN} and \ref{TL} agree with the specialization
to $D=4$ of Capper's results (\ref{flatsigma1}-\ref{T4}).

The advantages of our de Sitter gauge (\ref{dSgauge}) are so great that it has
been used for nine \cite{Tsamis:1996qk,Tsamis:2005je,Miao:2005am,Kahya:2007bc,
Miao:2012bj,Leonard:2013xsa,Boran:2014xpa,Boran:2017fsx,Glavan:2020gal} of the 
ten graviton loops which have so far been computed de Sitter background. The 
exception was a year-long {\it tour de force} made to check for gauge dependence 
in the vacuum polarization \cite{Glavan:2015ura} using a cumbersome, 1-parameter
family of de Sitter invariant gauges \cite{Mora:2012zi}. It would be quite 
challenging re-computing the graviton self-energy in this family of gauges. We
have instead devised a 2-parameter deformation of the de Sitter breaking gauge
(\ref{dSgauge}) \cite{Glavan:2019msf},
\begin{equation}
\mathcal{L}^{\alpha\beta}_{GF} = -\frac{a^{D-2}}{2 \alpha} \eta^{\mu\nu} F_{\mu} 
F_{\nu} \;\; , \;\; F_{\mu} = \eta^{\rho\sigma} \Bigl( h_{\mu\rho , \sigma} - 
\frac{\beta}{2} h_{\rho\sigma , \mu} + (D\!-\! 2) a H h_{\mu\rho} 
\delta^{0}_{~\sigma} \Bigr) \; . \label{2param}
\end{equation}
Although we have not yet computed the graviton self-energy in this gauge, Capper
derived a result for its flat space limit \cite{Capper:1979ej}. The final, 
renormalized result takes the same form as (\ref{flatsigma3}) but with the
numerical coefficients changed. In the general gauge the coefficient of 
$\Pi^{\mu\nu} \Pi^{\rho\sigma}$ becomes \cite{Capper:1979ej},
\begin{equation}
\frac{23}{2} \longrightarrow \frac{45}{2} \alpha^2 \!+\! \frac{113}{2} \!+\!
\frac{15}{2} \frac{(\alpha \!-\! 3)^2}{(\beta \!-\! 2)^4} \!-\! \frac{135}{2}
\frac{(\alpha \!-\! 3)}{(\beta \!-\! 2)^3} \!-\! \frac{25}{2} \frac{(2 \alpha 
\!-\! 19)}{(\beta \!-\! 2)^2} \!+\! \frac{5}{2} \frac{(11 \alpha \!+\! 59)}{
(\beta \!-\! 2)} \; . \label{Coef1}
\end{equation}
The coefficient of $\Pi^{\mu (\rho} \Pi^{\sigma) \nu}$ becomes 
\cite{Capper:1979ej},
\begin{equation}
\frac{61}{2} \longrightarrow \frac{15}{2} \alpha^2 \!+\! \frac{45}{4} \alpha 
\!-\! \frac{43}{4} \!+\! \frac{5}{2} \frac{(\alpha \!-\! 3)^2}{(\beta \!-\! 2)^4} 
\!-\! \frac{105}{4} \frac{(\alpha \!-\! 3)}{(\beta \!-\! 2)^3} \!-\! \frac{5}{4} 
\frac{(\alpha \!-\! 51)}{(\beta \!-\! 2)^2} \!+\! \frac{5}{4} \frac{(9 \alpha 
\!-\! 11)}{(\beta \!-\! 2)} \; . \label{Coef2}
\end{equation}
Expressions (\ref{Coef1}) and (\ref{Coef2}) can be made arbitrarily positive by
taking $\beta$ near 2. They do seem to be bounded below, but they can definitely
change sign, and there are two real solutions which cause them both two vanish,
\begin{eqnarray}
\alpha \simeq 0.551886 \qquad & , & \qquad \beta \simeq 1.42999 \; , \\
\alpha \simeq 2.24351 \qquad & , & \qquad \beta \simeq 1.78159 \; .
\end{eqnarray}

The gauge dependence we have exhibited in the flat space results 
(\ref{Coef1}-\ref{Coef2}) must of course be present in the de Sitter result.
However, it still is not clear what happens to the parts of $-i [\mbox{}^{\mu\nu}
\Sigma^{\rho\sigma}](x;x')$ which represent the effects of inflationary gravitons.
To understand this better, consider the contributions to $T^{19}(x;x')$ from 
Tables~\ref{TN} and \ref{TL},
\begin{eqnarray}
\lefteqn{ T^{19}(x;x') = -\frac{\kappa^2}{64 \pi^4} \Biggl\{ -\frac{\frac{112}{5}
\Delta \eta}{\Delta x^6} + \frac{(\frac{26}{3} a \!+\! 8 a') H}{\Delta x^4} -
\frac{a^2 {a'}^2 H^4 \Delta \eta}{\Delta x^2} } \nonumber \\
& & \hspace{3.5cm} + \Biggl[ \frac{8 a a' H^2 \Delta \eta}{\Delta x^4} + 
\frac{2 a a' (a \!-\! 3 a') H^3}{\Delta x^2} \Biggr] \ln(H^2 \Delta x^2) \Biggr\}
\; . \qquad \label{T19full}
\end{eqnarray}
The flat space result consists of just the first term. It is this term whose
coefficient can be driven to infinity, or made to vanish by the gauge dependence
of (\ref{Coef1}-\ref{Coef2}). This term has no effect on the graviton mode function,
and induces fractional corrections to the potentials of the form $\kappa^2/r^2$.
How the parameters $\alpha$ and $\beta$ of the general de Sitter gauge (\ref{2param})
affect the other terms is not known. These other terms can potentially change the
graviton mode function, and they typically induce fractional changes in the potentials
of the form $G H^2$ times large temporal and/or spatial logarithms.

In flat space Donoghue and collaborators have shown how to extract unique, gauge
independent results for the fractional $\kappa^2/r^2$ correction to the potentials
\cite{BjerrumBohr:2002ks,BjerrumBohr:2002kt}. Their technique \cite{Donoghue:1993eb,
Donoghue:1994dn} is to first compute the one loop scattering amplitude between two 
massive particles, then use inverse scattering theory to infer the exchange potential. 
It was recently discovered that this process can be short-circuited in order to 
directly purge the vacuum polarization of gauge dependence \cite{Miao:2017feh}. The 
procedure is to assemble the same diagrams whose sum would produce the scattering 
amplitude, however, one works in position space and employs a series of identities 
which permit the higher point diagrams to be viewed as corrections to the 1PI 2-point 
function. For example, one of the many diagrams which contribute to the scattering of 
two massive scalars consists of two graviton lines emerging from the vertex at 
${x'}^{\mu}$ and attaching to the other massive scalar at points $x^{\mu}$ and 
$y^{\mu}$. This diagram does not have the 2-point topology to be viewed as a 
contribution to the graviton self-energy, however, Donoghue and collaborators have 
derived a series of reductions that capture the nonanlytic parts of the full 
amplitude which are responsible for infrared phenomena \cite{Donoghue:1993eb,
Donoghue:1994dn,Donoghue:1996mt,BjerrumBohr:2002sx}. If $i\Delta_m(x;x')$ denotes 
the massive scalar propagator then the relevant Donoghue Identity for the 3-point 
contribution just described is \cite{Miao:2017feh},
\begin{equation}
i\Delta_m(x;y) i\Delta(x;x') i\Delta(y,x') \longrightarrow 
\frac{i \delta^D(x\!-\!y)}{2 m^2} \Bigl[ i\Delta(x;x')\Bigr]^2 \; . 
\label{DonoID}
\end{equation}
Applying (\ref{DonoID}) reduces the 3-point contribution to a 2-point form which
can be viewed as a correction to the graviton self-energy. When all such corrections
are combined, dependence upon $\alpha$ and $\beta$ drops out and one is left with
a unique and gauge independent result \cite{Miao:2017feh}.

\section{Epilogue}

Quantum corrections from inflationary gravitons \cite{Miao:2005am,Kahya:2007bc,
Miao:2012bj,Leonard:2013xsa,Boran:2014xpa,Glavan:2015ura,Boran:2017fsx,Glavan:2020gal} 
modify how other particles propagate \cite{Miao:2006gj,Kahya:2007cm,Wang:2014tza,
Glavan:2016bvp,Boran:2017cfj,Glavan:2020ccz}, and the force laws they mediate 
\cite{Glavan:2013jca}. At one loop order these results involve a single graviton 
propagator, and it is principally the ``tail'' part of this propagator that 
engenders the most interesting effects \cite{Miao:2008sp,Miao:2018bol}. Quantum 
gravity corrections to gravity itself are even more interesting because they involve 
{\it two} graviton propagators at one loop order. The potential for gravity to 
mediate more interesting effects than matter can be seen from the factor of 
$\ln(H^2 \Delta x^2)$ which multiplies all the contributions of Table~\ref{TL}, 
and is absent from the analogous scalar contributions of Table~\ref{Tscalar}. 

The 1PI 2-graviton function $-i[\mbox{}^{\mu\nu} \Sigma^{\rho\sigma}](x;x')$ 
quantifies corrections to linearized gravity from matter and from gravity itself.
The point of this paper has been to develop a representation for the tensor 
structure of this object in terms of differential operators acting on structure 
functions. Matter contributions must be annihilated when the Ward operator 
(\ref{Wardop}) acts on {\it either} coordinate, but gravity contributions are only 
annihilated when the Ward operator acts on {\it both} coordinates. On flat space 
background one requires two structure functions for matter contributions and three 
for contributions from gravity, as in expression (\ref{flatsigma2}). The absence of 
time translation invariance and Lorentz invariance in cosmology means that four 
structure functions are required for matter contributions whereas nine are needed 
for contributions from gravity. Our representations are given in expressions 
(\ref{mattersigma}) for matter, and (\ref{gravitonsigma}) for gravity.

Quantum field theory computations typically express $-i[\mbox{}^{\mu\nu} 
\Sigma^{\rho\sigma}](x;x')$ as a linear combination of basis tensors which do not
individually obey the relevant Ward identity. For example, the flat space result
(\ref{flatsigma1}) was originally reported \cite{Capper:1979ej} using a basis of 
five tensors, which can then be organized into three combinations (\ref{flatsigma2})
that obey the Ward identity. This procedure for passing from raw results to
structure functions is known as {\it reconstruction}. Our reconstruction procedure
for cosmology is based on first recasting the primitive result as a sum 
(\ref{initialrep}) of the 21 tensor differential operators $[\mbox{}^{\mu\nu} 
\mathcal{D}^{\rho\sigma}]$ listed in Table~\ref{Tbasis}, each acting on a scalar
coefficient function $T^i(x;x')$. The $[\mbox{}^{\mu\nu} \mathcal{D}^{\rho\sigma}]$
are constructed from $\delta^{\mu}_{~0}$ and the spatial parts of the Minkowski 
metric $\overline{\eta}^{\mu\nu} \equiv \eta^{\mu\nu} + \delta^{\mu}_{~0} 
\delta^{\nu}_{~0}$ and the derivative operator $\overline{\partial}^{\mu} \equiv
\partial^{\mu} + \delta^{\mu}_{~0} \partial_0$. A typical example is furnished by
the scalar contribution (\ref{scalar3pt}) on de Sitter background  
\cite{Park:2011ww,Park:2015kua}. We first $3+1$ decompose derivatives 
(\ref{D1}-\ref{D2}) of the de Sitter length function, then express factors of
the spatial coordinate interval $\vec{x} - \vec{x}'$ as gradients using relations
(\ref{Rule1}-\ref{Rule3}). Our fundamental structure functions for matter are 
$T^{12}(x;x')$, $T^{16}(x;x')$, $T^{18}(x;x')$ and $T^{19}(x;x')$. For gravity
we express one action of the Ward operator in the form (\ref{selfmatter}), 
involving the ten basis tensors of Table~\ref{Sbasis} acting on coefficient
functions $S^i(x;x)$. The expansion of each $S^i$ in terms of the $T^i$ is given
in Table~\ref{SExpansion}. Because acting the Ward operator a second time must 
produce zero, the ten $S^i$ obeys the five relations given in Table~\ref{RExpansion}.
We take the five new structure functions for gravity to be $S^2(x;x')$, $S^4(x;x')$,
$S^7(x;x')$, $S^8(x;x')$ and $S^{10}(x;x')$. We have found it convenient to group
some of the fundamental structure functions ($T^{16}$, $T^{19}$, $S^4$, $S^8$ and
$S^{10}$) into two symmetric auxiliary functions, $\alpha(x;x')$ and $\beta(x;x')$, 
which are defined in expressions (\ref{gammadef}-\ref{betadef}). Our final 
representations for the self-energy in terms of the fundamental structure functions 
are expressions (\ref{mattersigma}) and (\ref{gravitonsigma}).

The formalism we have derived for representing the graviton self-energy improves 
on previous results \cite{Park:2011ww,Leonard:2014zua} in three ways:
\begin{itemize}
\item{It applies for contributions from gravitons in addition to contributions
from matter;}
\item{It is valid for any cosmological background (\ref{geometry}), not just
for de Sitter; and}
\item{Its structure functions involve the same functional forms as the primitive
result.} 
\end{itemize}
One can appreciate the final point from the explicit results for a loop of massless,
minimally coupled scalars \cite{Park:2011ww,Park:2015kua}. Primitive contributions
to the $T^i(x;x')$ consist of sums of products of non-negative powers of the two 
scale factors and the temporal separation $\Delta \eta$, times inverse powers of 
the Poincar\'e interval $\Delta x^2 \equiv (x - x')^2$. Because the fundamental
structure functions of this new representation are just $T^{12}$, $T^{16}$, 
$T^{18}$ and $T^{19}$, they of course have the same form. One might worry about the
auxiliary functions $\alpha$ and $\beta$, but expressions (\ref{T2final}) and 
(\ref{T7Final}) guarantee that they involve no new functional forms. Compare that
with what happens in the {\it simplest} of the previous representations. The 
renormalized spin zero structure function roughly equivalent to $T^{18}$ and a 
combination of $T^{16}$ and $T^{19}$ is \cite{Leonard:2014zua},
\begin{eqnarray}
\lefteqn{F_{0R} = \frac{\kappa^2}{9 (4\pi)^4} \Biggl\{ \frac{\partial^2}{2} 
\Biggl[ \frac{\ln(H^2 \Delta x^2)}{\Delta x^2}\Biggr] + a^2 {a'}^2 H^4 \Biggl[
-\frac{6}{y} \!+\! 6 } \nonumber \\
& & \hspace{4.5cm} \Bigl(-\frac{2}{y} \!+\! 6 \!-\! \frac{4}{4 \!-\! y} \Bigr)
\ln\Bigl( \frac{y}{4}\Bigr) + \frac32 (2 \!-\! y) \Psi(y)\Biggr\} , \qquad
\end{eqnarray}
where $y \equiv a a' H^2 \Delta x^2$ is the de Sitter length function and we
define 
\begin{equation}
\Psi(y) \equiv \frac12 \ln^2\Bigl( \frac{y}{4}\Bigr) - \ln\Bigl(1 \!-\! \frac{y}{4}
\Bigr) \ln\Bigl( \frac{y}{4}\Bigr) - {\rm Li}_2\Bigl( \frac{y}{4}\Bigr) \; ,
\end{equation}
where ${\rm Li}_{2}(x) \equiv -\int_{0}^{x} dt \ln(1-t)/t$ is the dilogarithm function.
The same exotic functional forms appear in the two tensor structure functions,
$F_{2R}(x;x')$ and $G_{2R}(x;x')$, which are roughly equivalent to $T^{12}(x;x')$ 
and a different combination of $T^{16}$ and $T^{19}(x;x')$. Deriving these 
structure functions from the primitive result for $-i[\mbox{}^{\mu\nu} 
\Sigma^{\rho\sigma}](x;x')$ is a major undertaking because it entails solving 
partial differential equations. Those equations were barely tractable for the 
scalar contributions owing to the absence of de Sitter breaking, but they become 
hopelessly complicated when de Sitter invariance is lost with graviton 
contributions. Finally, it is of course difficult {\it using} the exotic structure 
functions to solve the effective field equations. The new formalism obviates all 
of these problems.

The point of devising this representation is to solve the effective field equations.
Section 3.3 specializes the effective field equations for a graviton contribution to 
the cases of a spatial plane wave graviton (\ref{modeeqn}) and the two scalar potentials 
(\ref{onepotA}-\ref{onepotB}) which represent the gravitational response to point mass.
Both of these things have already been computed (using the old formalism) for the 
contribution of a massless, minimally coupled scalar \cite{Park:2011ww,Park:2015kua}.
Although there are no changes in the graviton mode function \cite{Park:2011ww}, the 
response to a point mass acquires corrections which grow at late times and large 
distances \cite{Park:2015kua}, 
\begin{eqnarray}
\lefteqn{ \Psi(\eta,r) = -\frac{GM}{ar} \Biggl\{1 + \frac{G}{20\pi a^2 r^2} }
\nonumber \\
& & \hspace{2cm} + \frac{G H^2}{\pi} \Bigl[-\frac1{30} \ln(a) - \frac{3}{10} 
\ln(a H r) \Bigr] + O(G^2) \Biggr\} , \qquad \\
\lefteqn{\Phi(\eta,r) = -\frac{GM}{ar} \Biggl\{ 1 - \frac{G}{60 \pi a^2 r^2} }
\nonumber \\
& & \hspace{2cm} + \frac{G H^2}{\pi} \Bigl[-\frac1{30} \ln(a) - \frac3{10} 
\ln(a H r) + \frac23 a H r\Bigr] + O(G^2) \Biggr\} . \qquad
\end{eqnarray}
We are now in a position to study what gravity does to itself. Applying the Hartree 
approximation indicates that inflationary gravitons enhance the ``electric'' 
components of the Weyl field strength \cite{Mora:2013ypa},
\begin{equation}
C^{\rm 1~loop}_{0i0j}(\eta,k) \longrightarrow -\frac{8}{\pi} G H^2 \ln(a) \times
C^{\rm tree}_{0i0j}(\eta,k) \; .
\end{equation}
It would be very interesting to extend the $D=4$ results of Tables~\ref{TN}, 
\ref{TL}, \ref{SN} and \ref{SL} to recover fully renormalized results, and then
employ them to solve equations (\ref{modeeqn}) and (\ref{onepotA}-\ref{onepotB}). 

\vspace{.5cm}

\centerline{\bf Acknowledgements}

This work was partially supported by the European Union's Seventh 
Framework Programme (FP7-REGPOT-2012-2013-1) under grant agreement 
number 316165; by the European Union's Horizon 2020 Programme
under grant agreement 669288-SM-GRAV-ERC-2014-ADG;
by NSF grants PHY-1506513 and 1806218; and by the UF's Institute for 
Fundamental Theory.

\end{document}